\documentclass[a4paper,usenatbib]{mn2e}
\usepackage{graphicx}
\usepackage{psfrag}
\usepackage{amsmath}
\usepackage{amssymb}
\usepackage{color}
\usepackage{hyperref}
\hypersetup{
    colorlinks=true,
    citecolor=red,
    filecolor=black,
    linkcolor=blue,
    urlcolor=magenta,
    linktocpage=true,
    breaklinks = true
}

\usepackage{subfigure,float}
\usepackage{multirow}
\usepackage[normalem]{ulem}

\usepackage{array,booktabs}

\def\aj{Astron. J.}
\def\apj{Astrophys. J.}
\def\apjl{Astrophys. J.}
\def\mnras{Mon. Not. Roy. Astron. Soc.}
\def\prd{Phys. Rev.}

\def\aap{Astron. Astrophys.}
\def\apjs{Astrophys. J. Suppl.}

\def\physrep{Physics Reports}
\def \pasj {Publ. Astron. Soc. Jap.}
\def \pasa {Publ. Astron. Soc. Austral.}

\def \pasp {Publ. Astron. Soc. Pac}

\newcommand{\beq}{\begin{equation}}
\newcommand{\eeq}{\end{equation}}

\newcommand{\ber}{\begin{eqnarray}}
\newcommand{\eer}{\end{eqnarray}}

\def \dd {\rm d}
\def\rhi {\rho_{\rm HI}}
\def\ehi {\eta_{\rm HI}}

\def\ehic {\eta_{\rm HI}^C}

\def \dehic{\delta \eta_{\rm HI}^C}
\def\eth {\eta_{\rm HI}^{th}}
\def \dehi {\delta \eta_{\rm HI}}
\def \deth {\delta \eta_{\rm HI}^{th}}
\def \meanrh {\bar{\rho}_H}
\def \ffc {FF}
\def \meanrhi {\bar{\rho}_{\rm HI}}
\def \xhi {x_{\rm HI}}

\title[Morphology of HI density fields]{Studying the morphology of HI isodensity surfaces during reionization 
using Shapefinders and percolation analysis}

\author[Bag, Mondal, Sarkar, Bharadwaj, Choudhury, Sahni]{Satadru Bag,$^1$\thanks{E-mail: satadru@iucaa.in} Rajesh Mondal,$^{2}$ Prakash Sarkar,$^3$ 
\newauthor
Somnath Bharadwaj,$^4$ Tirthankar Roy Choudhury$^5$ and Varun Sahni$^1$
\\
$^{1}$ Inter-University Centre for Astronomy and Astrophysics, Pune, India\\
$^{2}$ Astronomy Centre, Department of Physics and Astronomy, University of Sussex, Brighton, BN19QH, UK\\
$^{3}$ National Institute of Technology, Jamshedpur, India\\
$^{4}$ Department of Physics and Centre for Theoretical Studies, Indian Institute of Technology Kharagpur, Kharagpur 721302, India\\
$^{5}$ National Centre for Radio Astrophysics, Pune, India
}

\date{}

\pubyear{2015}

\begin{document}
\label{firstpage}
\pagerange{\pageref{firstpage}--\pageref{lastpage}}
\maketitle

\begin{abstract}
Minkowski functionals and Shapefinders shed light on the connectedness of large-scale structure by determining its topology and morphology.
We use a sophisticated code, SURFGEN2, 
to measure the Minkowski functionals and Shapefinders of individual clusters by modelling cluster surfaces using the {\em Marching Cube 33} triangulation algorithm.
 In this paper, we study the morphology of simulated neutral hydrogen (HI) density fields using Shapefinders at various stages of reionization from the excursion set approach. 
Accompanying the Shapefinders, we also 
employ the `largest cluster statistic' (LCS) to understand the percolation process. {\em Percolation curves} demonstrate that the non-Gaussianity in the HI field increases as reionization progresses. 
The large clusters in both the HI overdense and underdense excursion sets possess similar values of ``thickness'' ($T$), as well as ``breadth'' ($B$),  but their third Shapefinder -- ``length'' ($L$) -- becomes almost proportional to their volume. The large clusters in both HI overdense and underdense segments are overwhelmingly filamentary.  The `cross-section' of a filamentary cluster can be estimated using the product of the first two Shapefinders, $T \times B$.  Hence the cross sections of the large clusters at the onset of percolation do not vary much with volume and their sizes only differ in terms of their lengths. 
This feature appears more vividly in HI overdense regions than in underdense regions and is more pronounced at lower redshifts 
which correspond to an advanced stage of reionization. 

\end{abstract}




\section{Introduction}
The epoch of reionization (EoR) is one of the important but perhaps the least
known phase in the evolution history of our universe. During this
epoch, the diffuse hydrogen in the intergalactic medium (IGM)
gradually changed its state from being neutral (HI) to ionized
(HII). Our knowledge about this epoch is guided so far by the
observations of the Thomson scattering optical depth of the CMB
photons \citep{komatsu2011,Ade:2015xua,planck2016b}, the observations
of the Ly-$\alpha$ absorption spectra of the high-redshift
quasars \citep{becker2001, fan2003, goto2011, becker2015} and the
luminosity function and clustering properties of high-redshift
Ly-$\alpha$ emitters \citep{trenti2010, ouchi2010, ota2017,
zheng2017}. These observations together suggest that this epoch
probably extended over a broad redshift range $6 \lesssim z \lesssim
15$ \citep{mitra11, mitra2015, robertson2015}. However, our understanding on most of the 
fundamental issues associated with this epoch, such as the properties
of ionizing sources, topology of the neutral hydrogen and the
morphology of ionized HII regions etc. at different stages of the
reionization remain uncertain till date. 

Observations of the redshifted HI 21-cm signal which provides
a direct window to the state of the hydrogen in the IGM have the
potential to probe this complex epoch. There is a considerable effort
underway to detect the EoR 21-cm signal using radio interferometry
e.g. GMRT \citep{paciga2013}, 
LOFAR \citep{haarlem2013,yatawatta2013},
MWA \citep{bowman2013,dillon2014},
PAPER \citep{parsons2014,ali2015,jacobs2015}.
Apart from these first generation radio interferometers, the detection
of this signal from EoR is one of the key science goals of the future
radio telescopes e.g. SKA \citep{mellema2013,koopmans2015},
HERA \citep{furlanetto2009, DeBoer2017}. 

Analyses of 21-cm signals are mainly based on traditional $N$-point correlation statistics. Beyond the simplest two-point function (power spectrum),  higher order correlations are quite non-trivial to calculate and sometimes they suffer from conceptual challenges. On the other hand, the Minkowski functionals (MFs) are extremely useful tools in quantitatively describing the morphology because, in principle, they contain information on all the higher order moments. The MFs were first introduced in cosmology by \cite{mecke}. Since then they have been extensively employed to study the morphology of the large scale structure of the universe and the cosmic web \citep{Schmalzing1997, Sahni:1998cr, Sathyaprakash:1998gy, Bharadwaj2000, Hikage2003, Bharadwaj2004, Pandey2008, Einasto2011, Wiegand2017} as well as the CMB \citep{Schmalzing1998,Novikov1999,Novikov2000,Hikage2006}. Since the reionization landscape is similarly rich in geometrical properties because of growth and overlap of ionized ``bubbles'', studying the morphology of reionization using MFs is highly compelling and feasible. The physics underlying the reionization process is expected to be manifested in the
geometry and morphology of HI and HII regions. The ratios of MFs are introduced in \cite{Sahni:1998cr} as {\em Shapefinders} which precisely assess the shape of an object by directly estimating its physical dimensions. Therefore, using MFs and Shapefinders of the ionization field, it should be possible to probe the physics of the high-redshift universe. For instance, if reionization is driven in a
non-standard manner through large energy output from multiple quasar jets, the very first ionized
bubbles might be filamentary and not spherical (as they would be for point-like sources like stars in galaxies). Clearly the three
dimensional structure of cosmological reionization in this scenario would be very different
from the standard mechanism of point-like sources.             

In the past decade a number of efforts have been made to study the morphology of reionization including the study of the Minkowski functionals \citep{Gleser:2006su, Lee:2008, Friedrich2011,yoshiura:2017,Kapahtia2017, Bag:2018zon} as well as the percolation analyses \citep{Iliev2006,Iliev2014,Furlanetto2016}.  In the preceding work \citep{Bag:2018zon}, we studied the percolation process in the ionized segment in HI density fields and studied the size and shape distributions of ionized regions using Minkowski functionals and Shapefinder at various stages of reionization. We showed that the largest region in ionized hydrogen possesses a characteristic cross-section of $\sim 7$ Mpc$^2$ which remains almost constant across the percolation transition while its length increases abruptly, making it highly filamentary at the onset of percolation. As a continuation, in this second of a series of papers, we study the Shapefinders of clusters in both HI overdense and underdense segments from the excursion set approach. In conjunction with Shapefinders we also employ the percolation analysis\footnote{In the literature, percolation process has been studied comprehensively in the context of mathematical and condensed matter physics \citep{essam,Isichenko,stauffer,saberi}} using `largest cluster statistics' \citep{Klypin1993} in both the segments.  
Also the study of non-Gaussianity in HI density field or 21-cm signal has become very important over the time and drawn much attention recently \citep{Bharadwaj2005, Cooray2005, Pillepich2007, mondal2015, Yoshiura:2014ria, Bandyopadhyay2017, majumdar2017}. One concomitant advantage of percolation analysis is that the so called {\em percolation curves}, introduced in \cite{Sahni:1996mb} exhibit the asymmetry in percolation between the overdense and underdense excursion sets and provides a measure of the non-Gaussianity in the HI density field in geometrical point of view. Since percolation analysis is carried out in real space, this viable technique of estimating non-Gaussianity complements the tradition methods, such as bispectrum performed in Fourier space.  

We developed an advanced code, named SURFGEN2, which models the surfaces of clusters in 3-dimensions using {\em Marching Cube 33} triangulation algorithm \citep{mar33} and subsequently calculates their Minkowski functionals and associated Shapefinders. We compute MFs and Shapefinders of each individual clusters in both the HI overdense and underdense segments.
Our method is the much-improved and refined version of the surface modeling scheme, SURFGEN \citep{Sheth:2002rf}, which was employed to explore geometry and topology of the large-scale structure of the universe and the cosmic web \citep{Sheth2004,Shandarin:2003tx,Sheth2005, Einasto2007}. Our methodologies are empowered with the following advantages over the similar attempts in the literature. Firstly, Shapefinders directly estimate the extensions of each cluster in 3 dimensions, hence together with MFs they completely provide the physical shape and size, i.e. the geometry, morphology and topology, of each cluster. Secondly, SURFGEN2 models the cluster surfaces through advanced {\em Marching cube 33} triangulation algorithm resulting in much better accuracy in computing the MFs and Shapefinders compared to the existing methods \citep{Schmalzing1997}, such as Crofton's formula \citep{Crofton_1968} or Koenderink's invariant \citep{Koenderink}. Additionally, the novel approach for estimating non-Gaussianity using percolation curves avoids many difficulties associated with conventional methods carried out in Fourier space.

Our paper is organized as follows. The Minkowski functionals and Shapefinders are briefly described in section \ref{sec:Minkowski}. In section \ref{sec:SURFGEN2}, the methodologies  of the code SURFGEN2 are briefly discussed. Section \ref{sec:data} explains the procedure of simulating the HI density field. The cluster statistics in both HI overdense and HI underdense segments are studied in section \ref{sec:cs} from excursion set approach. Section \ref{sec:shape} contains the analyses of shapes of clusters in both overdense and underdense regions. The conclusions are presented in section \ref{sec:conclusion}.

\section{Minkowski functionals and Shapefinders}\label{sec:Minkowski}

The morphology of a closed two dimensional  surface embedded in three dimensions 
is well described by the four Minkowski functionals \citep{mecke}
\begin{enumerate}
 \item Volume enclosed: $V$,
 
 \item Surface area: $S$,
 
 \item Integrated mean curvature (IMC):
 \begin{equation}
  C=\frac{1}{2} \oint \left(\frac{1}{R_1}+\frac{1}{R_2} \right) dS \;,
 \end{equation}
 
  \item Integrated Gaussian curvature or Euler characteristic: 
  \begin{equation}
  \chi=\frac{1}{2\pi} \oint \frac{1}{R_1 R_2} dS\;.
  \end{equation} 
\end{enumerate}

Here $R_1$ and $R_2$ are the two principal radii of curvature at any point on the surface. 
The fourth Minkowski functional (Euler characteristic) can be written in terms of the genus (G) 
of the surface as follows,
\begin{equation}
 G=1-\chi/2 \equiv {\rm (no.~ of~ tunnels)}-{\rm (no.~ of~ isolated~ surfaces)}+1\;.
\end{equation}
It is well known that $\chi$ (equivalently $G$) is a  measure of the topology of the surface.

The `Shapefinders', introduced in \cite{Sahni:1998cr}, are ratios of these Minkowski functionals,
namely
\begin{align}
 {\rm Thickness:}~T &=3V/S\;, \\
 {\rm Breadth:}~B &=S/C\;, \\
 {\rm Length: }~L&=C/(4\pi)\;. \label{eq:L}
\end{align}

The Shapefinders -- $T, B, L$ -- have dimension of length, and estimate the three physical extensions of an object in 3-dimensions\footnote{In general one finds $L \geq B \geq T$. However, if the natural order $T\leqslant B \leqslant L$ is not maintained, we choose the smallest dimension as $T$ and the largest one as $L$ to restore the order. In the rare cases when a cluster has 
$C < 0$ we shall redefine $C \to |C|$ to ensure that $B$ and $L$ are positive.}.
The Shapefinders are spherically normalized, i.e. $V=(4\pi/3) TBL $. 
 
Using the Shapefinders one can determine the morphology of an object (such as
an isodensity surface),
by means of the following dimensionless quantities\footnote{One can redefine `Length' by taking the genus of an object into account \cite{Sheth:2002rf},
\begin{equation}\label{eq:L1_HI}
L_1=\frac{C}{4 \pi (1+|G|)}\;. \nonumber
\end{equation}
 This reduces the filamentarity in the following manner while keeping planarity unchanged,
 \begin{equation}\label{eq:f1_HI}
  F_1=\frac{L_1-B}{L_1+B}\;. \nonumber
 \end{equation}
 These definitions, $L_1$ and $F_1$, receptively assess the `macroscopic' length and filamentarity of a given object. On the other hand, the definitions of length and filamentarity, given by \eqref{eq:L} and \eqref{eq:F} respectively, provide us with the microscopic information which we are interested in.}
 which characterize
 its planarity and filamentarity \citep{Sahni:1998cr}
\begin{align}
{\rm Planarity:}~ P&=\frac{B-T}{B+T}\;,\label{eq:P} ~~\\ 
{\rm Filamentarity:}~ F&=\frac{L-B}{L+B}\;. \label{eq:F}
\end{align}
For a planar object (such as a sheet) $P \gg F$, while the reverse is true for a filament which has
$F \gg P$. A ribbon will have $P \sim F$ whereas $P \simeq F \simeq 0$ for a sphere.
In all cases $ 0 \leq P,F \leq 1$. Therefore, Shapefinders, together with Minkowski functionals, provide us with all the information about the geometry, morphology and topology of a 3-dimensional field.

\section{The code SURFGEN2} \label{sec:SURFGEN2}
In this section we briefly discuss the code SURFGEN2 which
constructs isodensity surfaces from a given density field and subsequently determines their 
morphology. 
SURFGEN2 is a more advanced version of SURFGEN algorithm originally proposed by \cite{Sheth:2002rf, Sheth:2006qz}.
It consists of three parts.
\begin{itemize}

\item The first part identifies all clusters (overdense or underdense segments) 
within the simulation box using a `Friends-of-Friends' (FoF) algorithm with periodic boundary conditions. 
In the (rare) case when
fragmented parts of one and the same cluster are located at the box boundary these parts
 are rearranged to construct the cluster as demonstrated in figure 1 of \cite{Bag:2018zon}. 
 
 \item The second part of the code triangulates the surface of each cluster and stores the triangles' vertices. The triangulation method, using {\em `Marching Cube'} algorithm \citep{marcube}, is described in \cite{Sheth:2002rf, Sheth:2006qz} in detail. Instead, we use the improved triangulation scheme, known as `{\em Marching Cube 33}' \cite{mar33}, which circumvents the issues associated with the original Marching Cube algorithm. 
 
 \item The last part of the code determines the Minkowski functionals (and Shapefinders) for each 
cluster from the stored triangle vertices. The algorithms to calculate the Minkowski functionals have been briefly discussed below. 
\end{itemize}

\subsection{Determining the Minkowski functionals and Shapefinders}

Having triangulated an isodensity surface with the Marching Cube 33 algorithm
one can determine the Minkowski functionals and Shapefinders by the following means
\citep{Sheth:2002rf}, 
\begin{enumerate}
\item The {\bf surface area (S)} of the isodensity surface can be determined by summing over 
all the triangles
  constituting the surface
  \begin{equation}
    \label{eq:area}
    S = \sum_{i=1}^{N_T}S_i
  \end{equation}
  where $S_i$ is the area of the $i^{th}$ triangle and $N_T$ is the
  total number of triangles. 
\item The {\bf volume ($V$)} enclosed by the isodensity surface is given by the equation
  \begin{equation}
    \label{eq:vol}
    V=\sum_{i=1}^{N_T} V_i,\ \  V_i=\frac{1}{3}S_i({\bf \hat{n}}_i.{\bf P}_i)
  \end{equation}
  where $V_i$ is the volume of the $i^{th}$ tetrahedron whose base is the $i^{th}$ triangle and 
apex is (an arbitrarily chosen) origin. 
  ${\bf P}_i$ is the position vector of the centroid of the
  $i^{th}$ triangle having normal in ${\bf \hat{n}}_i$ direction. This
 method is explained in detail in \cite{Sheth:2002rf}.

  \item {\bf Integrated Mean Curvature (IMC)} is determined by the formula 
  \begin{equation}
    \label{eq:imc}
    C=\sum_{i,j}\frac{1}{2} \epsilon\  l_{ij}\phi_{ij}\;,
  \end{equation}
  where $l_{ij}$ and $\phi_{ij}$  are respectively the length of the common edge and the angle between the normals of the adjacent pair of triangles $i,j$. The summation is over all pairs of 
adjacent triangles.  $\epsilon$ takes the value $+1$ when the triangle pair $(i,j)$ is the part of a convex surface locally and $-1$ when the surface is concave locally. The detailed
 procedure for calculating the integrated mean curvature is explained in appendix \ref{appendix:imc}.
  
  The reader might note that
we also determine the IMC using the per-vertex method described in \cite{Rusinkiewicz2004}. 
However it has been our experience that for the most deformed surfaces, equation \eqref{eq:imc} provides a
 much better estimate for IMC than the per-vertex method. 
Therefore in this paper we primarily use \eqref{eq:imc} to calculate the IMC, 
utilizing the per-vertex method for an independent consistency check of our results.   
  
\item {\bf The Euler characteristic ($\chi$)} and the {\bf genus ($G$)} for a closed 
triangulated surface are
  given by the following expressions, 
  \begin{equation}
    \label{eq:G}
    \chi=N_T-N_E+N_V, \ \ G = 1-\frac{\chi}{2}
  \end{equation}
where $N_T$, $N_V$ and $N_E$ are the number of triangles, vertices and
edges respectively\footnote{For a closed surface, triangulated using {\rm Marching Cube 33} algorithm, each edge is always shared by two 
neighboring triangles. This leads to the relation $N_E=(3/2)N_T$ between the total number of
edges, $N_E$, and total number of triangles, $N_T$, comprising
 a closed surface. This relation can be used to check the topological
correctness of a given triangulation scheme.}. 
Since the above equation is just a portrayal of Euler's polyhedral formula, the Euler characteristic ($\chi$) calculated using this method is always exact.

\end{enumerate}
Since SURFGEN (and this advanced scheme SURFGEN2) models the surface of individual clusters through triangulation, the accuracy of SURFGEN is excellent, as demonstrated by \cite{Sheth:2002rf}, and much better than the existing methods of estimating the Minkowski functionals \citep{Schmalzing1997}, for example using the Koenderink invariant \citep{Koenderink} or the Crofton's formula \citep{Crofton_1968}.

\section{Simulating the neutral hydrogen density field}
\label{sec:data}
In this section, we briefly summarize the simulation of HI fields at
the different stages of the EoR. The reader is referred to section 2
of \citet{Bag:2018zon} for a detailed description. We have generated the
HI fields using semi-numerical simulations which closely
follow \citet{majumdar2014, mondal2015, mondal2016, mondal2017,
mondal2018} to simulate the ionization field. Our semi-numerical
method involves three following steps. 

In the first step, We have used a parallelized particle-mesh (PM)
$N$-body code to generate the dark matter distribution in a
$[215.04\,{\rm Mpc}]^3$ comoving box with a $3072^3$ grid using
$1536^3$ dark matter particles. We have run our simulation with
$0.07\,{\rm Mpc}$ spatial resolution which corresponds to a mass
resolution of $1.09\times 10^9\,M_{\odot}$.

In the next step, the standard friends-of-friends (FoF)
algorithm \citep{davis1985} was used to identify the location and mass 
of the collapsed halos in the dark matter distribution. We use a fixed
linking length 0.2 times the mean inter-particle separation, and
require a halo to have at least 10 particles which engenders the
minimum halo mass of the star-forming halos $M_{\rm halo,\,min}$. 

The final step generates the ionization map based on the
excursion set formalism of \citet{furlanetto2004} using the
homogeneous recombination scheme of \citet{Choudhury2009}. The
assumption here is that the hydrogen exactly traces the underlying dark
matter field and the halos host the ionizing sources. It is also
assumed that the number of ionizing photons ($N_{\gamma}$) emitted by
a source is proportional to the mass of the host halo ($M_{\rm halo}$)  
\begin{equation}
N_{\gamma}=N_{\rm ion} \frac{M_{\rm halo}}{m_{\rm H}}\,,
\label{nion}
\end{equation}
where the constant of proportionality $N_{\rm ion}$ is dimensionless
parameter and $m_{\rm H}$ is the hydrogen mass. In addition to minimum
halo mass $M_{\rm halo,\,min}$ and $N_{\rm ion}$, the simulations have
another free parameter $R_{\rm mfp}$, the mean free path of the
ionizing photons. The step described in this paragraph used a
low-resolution grid $8$ times coarser than the $N$-body simulations
i.e. grid spacing of $0.56\,{\rm Mpc}$. 

The redshift evolution of the neutral fraction
$\xhi(z)$ during the EoR is largely unknown. Given the uncertainty of
reionization history, we choose a fiducial model with the parameters
values $M_{\rm halo,\,min}=1.09\times 10^9\,M_{\odot}$, $N_{\rm
ion}=23.21$ and $R_{\rm mfp}=20\,{\rm Mpc}$ \citep{songaila2010} so as
to achieve 50\% ionization by $z=8$. The Thomson scattering optical depth $\tau=0.057$ for our reionization history. We have considered seven
different redshifts $z=[7,\,7.5,\,8,\,9,\,10,\,11,\,13]$ at which the
HI fields were generated. Note that, for simplicity, we assume $T_S \gg T_{\gamma}$, where $T_S$ and $T_{\gamma}$ are spin and CMB temperatures respectively, in our simulations.

It is convenient to define the `21-cm radiation efficiency' as \citep{Madau:1996cs, Bharadwaj:2004it}
\begin{equation}\label{eq:eta}
 \ehi({\bf x},z) \equiv \frac{\rhi({\bf x},z)}{\meanrh(z)}\;,
\end{equation}
where $\rhi$ and $\meanrh$ are neutral hydrogen density and mean hydrogen density respectively. Here we have assumed that $T_S \gg T_{\gamma}$. Therefore, the dimensionless quantity $\ehi({\bf x},z)$ can be regarded as a scaled neutral hydrogen density field in the comoving scale. Ignoring the redshift-space distortion, the 21-cm brightness temperature fluctuation is proportional to the dimensionless quantity 
\begin{equation} \label{eq:dehi}
\delta \eta_{\rm HI}({\bf x},z) \equiv \frac{\rho_{\rm HI}({\bf x},z)-\bar{\rho}_{\rm HI}(z)}{\bar{\rho}_H (z)}=\ehi({\bf x},z)-\xhi(z)\;,
\end{equation}
where the neutral fraction $\xhi(z) \equiv \meanrhi(z)/\meanrh(z)$.
$\dehi$ would be directly related to the observed quantity in EoR 21-cm experiments (such as Square Kilometre Array (SKA)). In this paper, we work with $\ehi$ and $\dehi$ instead of using the HI density $\rhi$.

\section{Cluster statistics and Percolation analysis}\label{sec:cs}
\begin{figure*}
\centering
\subfigure[overdense segment]{
\includegraphics[width=0.483\textwidth]{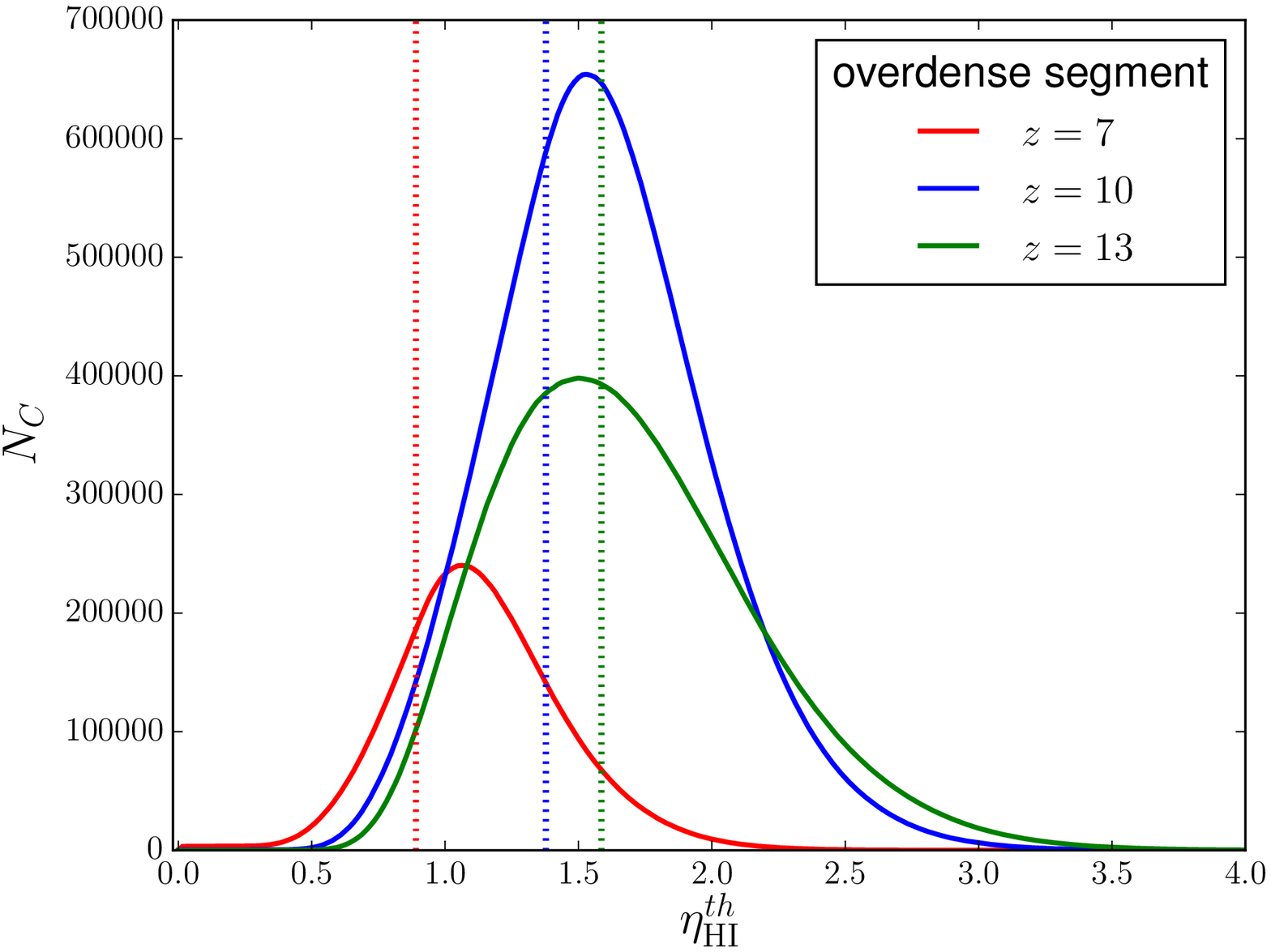}\label{fig:NC_vs_rhoth_struc}}
\subfigure[underdense segment]{
\includegraphics[width=0.483\textwidth]{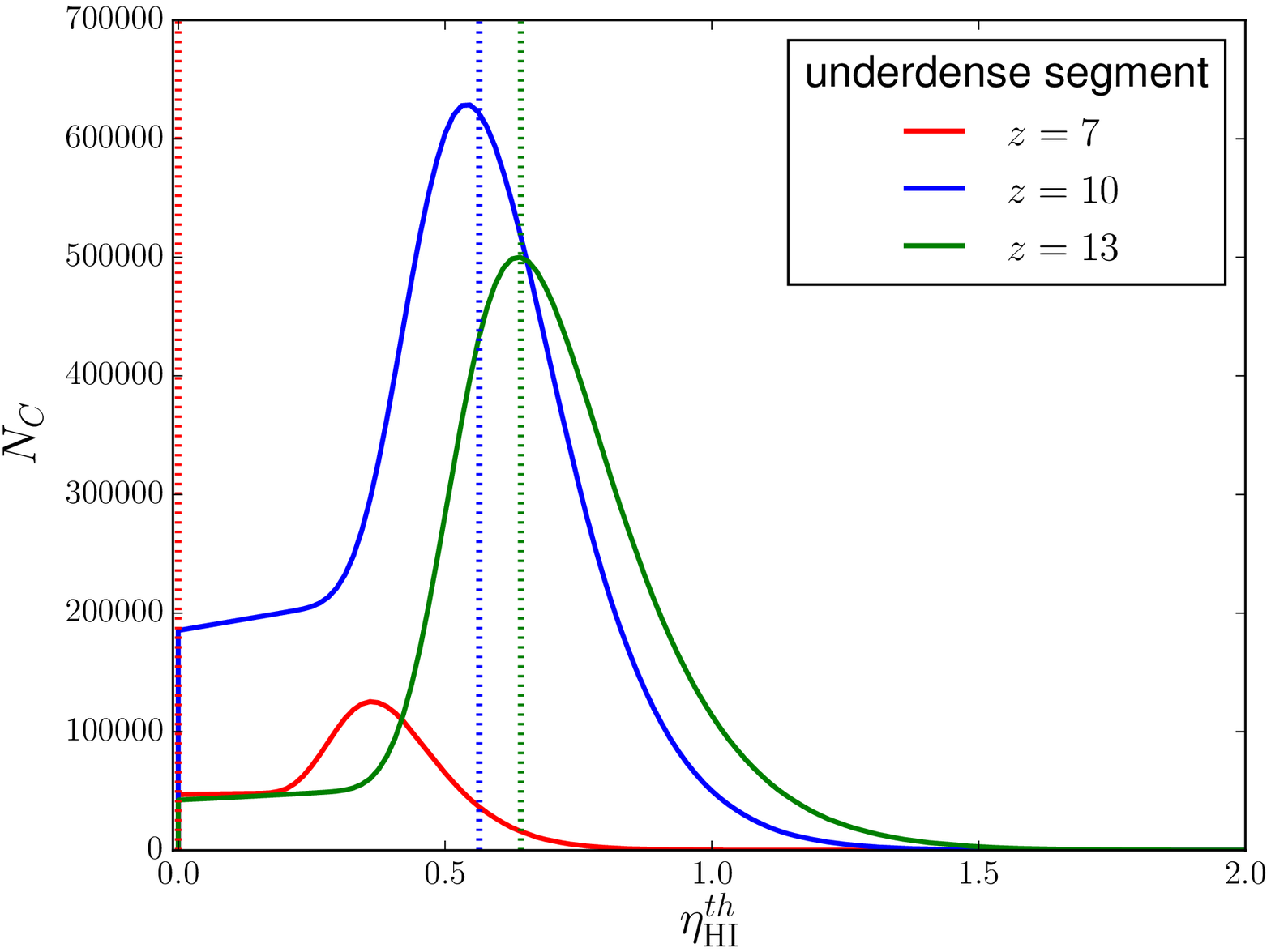}\label{fig:NC_vs_rhoth_void}}
\caption{The number of clusters ($N_C$) are shown as a function of the density threshold ($\eth$) at different redshifts for HI overdense (left panel) and underdense (right panel) segments. 
The percolation transitions in HI overdense and underdense segments are shown by the dotted vertical lines in the respective panels. The dotted lines use the same colour scheme as the solid lines to represent different redshifts.
 The discontinuous jump in $N_C$ of underdense segment near $\eth=0$, shown in the right panel, is because of the significant number of the completely ionized regions with $\rhi=0=\ehi$.  Although the volume of the completely ionized regions is higher at lower redshift, these regions are more interconnected at lower redshift. Therefore, the number of underdense clusters at $z=7$ near $\eth=0$ is actually smaller than that at $z=10$.    
 }
\label{fig:NC_rhoth_FF}
\end{figure*}

\begin{figure}
\centering
\includegraphics[width=0.47\textwidth]{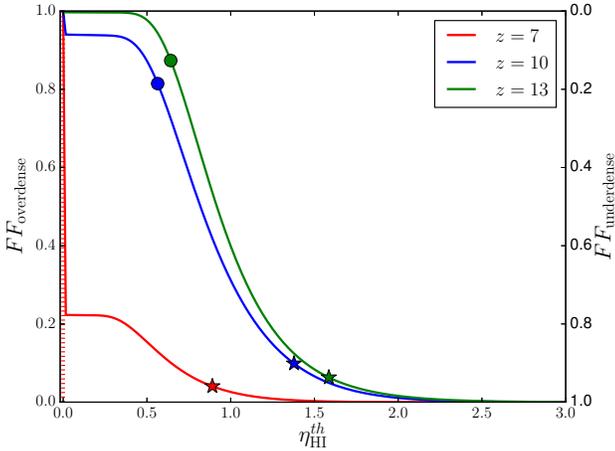}
\caption{ The filling factor ($\ffc$) vs $\eth$ curves at different redshifts are shown for HI overdense and underdense regions along the left y-axis and the right y-axis respectively. The underdense filling factor is given by $\ffc_{\rm underdense}=1-\ffc_{\rm overdense}$ always. The percolation transitions in overdense and underdense segments are marked on each curve by the star and the filled circle receptively. At any redshift, $\ffc$ of underdense segment sharply rises from zero near the threshold $\eth=0$ because of the presence of significant completely ionized regions with $\ehi=0$. This discontinuity is larger at smaller redshift owing to the fact that the completely ionized regions grows with reionization. The completely ionized regions percolate at $z \lesssim 9$, as explained in \protect\cite{Bag:2018zon}. So the critical filling factor at the percolation transition in HI underdense segment is not defined at $z=7$. Rather, the percolation transition in the underdense segment at $z=7$ is shown by the red dotted vertical line.}
\label{fig:FF_vs_rhoth}
\end{figure}

We first construct the isodensity surface, corresponding to a density threshold ($\eth$, $\ehi$ is defined in \eqref{eq:eta}), in the HI density fields such that the surface separates the HI over-density and under-density regions. For a chosen value of the threshold $\eth$, the two segments are defined by the following excursion sets\footnote{In the inside-out model of reionization, the high-density regions (of dark matter and hydrogen) are ionized first. Hence a region can become underdense in HI (subject to a threshold) due to either ionization or being void originally. On the other hand, HI overdense regions would not be originally very highly dense to host the ionizing sources.},  
\begin{align}\label{eq:segments}
&{\rm HI ~overdense ~segment~(21 ~cm ~hot ~spots):}~\ehi \geq \eth\;, \\ \nonumber
&{\rm HI ~underdense ~segment~(21 ~cm ~cold ~spots):}~\ehi < \eth\;.
\end{align}
Note that since the neutral hydrogen density is proportional to the brightness temperature, the overdense and underdense regions would correspond to the hot and cold spots respectively in the 21 cm observations with future radio surveys. We study both the HI overdense and the underdense excursion sets with varying $\eth$. We define `clusters' in each set separately with connected grid points using `Friends-of-Friends' (FoF) algorithm compatible with periodic boundary condition, as described in figure 1 of \cite{Bag:2018zon}.

The filling factor ($FF$) is defined for the HI overdense or the underdense segment as \footnote{
In section \ref{sec:cs}, we estimate the volume of individual HI overdense or underdense clusters
by counting the grid points inside each cluster. The number
of grid points inside a cluster roughly reflects its volume. On the other hand, one could calculate the volume of all clusters precisely through triangulation (as we perform in the next section), but the computation time would have been enormous.}, 
\begin{equation}\label{eq:ffc}
 FF=\frac{\rm total ~volume ~of ~all ~the ~clusters}{\rm volume ~of ~the ~simulation ~box}\;.
\end{equation}

In the figure \ref{fig:NC_rhoth_FF}, the number of cluster ($N_C$) are plotted against density threshold ($\eth$) for both HI overdense (left panel) and underdense (right panel) segments at different redshifts, $z=7,10$ and $13$. The corresponding filling factor vs $\eth$ curves are shown in figure \ref{fig:FF_vs_rhoth}. The filling factor of HI overdense and underdense excursion sets are plotted along the left y-axis and the right y-axis respectively. Note that $FF_{\rm overdense}+FF_{\rm underdense}=1$ in all the cases. For each cluster (of the overdense/underdense segment) we have  tracked whether the cluster extends from  one face of the box to the opposite face.  We consider a particular segment to have undergone the percolation transition when there exists at least one cluster which satisfies this condition.

The nature of the curves in figures \ref{fig:NC_rhoth_FF} and \ref{fig:FF_vs_rhoth} are expected \citep{Yess1996} and explained below for HI overdense and underdense regions separately. 
\begin{itemize}
 \item {\bf For HI overdense segment:} 
 At any redshift, there are very few highly HI dense regions and they are also small in size. Hence, the number of cluster ($N_C$) and the filling factor ($\ffc$) of HI overdense segment are very small when the density threshold ($\eth$) is very high. With the progress of reionization, the dense regions around ionization sources, located at the HI density peaks, are ionized first. Therefore, both $N_C$ and $FF_{\rm overdense}$ are smaller at lower redshift for high values of $\eth$, as evident from the figures \ref{fig:NC_vs_rhoth_struc} and \ref{fig:FF_vs_rhoth} respectively. As we lower $\eth$, the overdense clusters grow in size and as well as in number. Therefore, both $N_C$ and $\ffc$ at any redshift increase as we decrease $\eth$ from a very high value. As we keep decreasing $\eth$, eventually the overdense clusters start to merge vigorously. Consequently, the number of cluster starts to decrease but $\ffc$ continues to increase because the overdense segment keeps growing with decreasing $\eth$. Around this threshold, the largest cluster grows so rapidly that soon it extends from one face of the simulation box to the opposite face. Due to periodic boundary condition, such a cluster is formally infinite in extent and cannot be bound within the simulation box. This `phase transition' is referred to as the `percolation transition' which occurs at a critical threshold near the turnaround in the $N_C$ curve. Percolation transition in HI overdense segment at different redshifts has been shown by the vertical dotted lines in figure \ref{fig:NC_vs_rhoth_struc} and by the filled stars on the curves in figure \ref{fig:FF_vs_rhoth}.

 \item {\bf For HI underdense segment:} At all the redshifts, the number of clusters ($N_C$) in underdense excursion set varies with $\eth$ in exactly opposite way when compared to that of the overdense segment. Since the region with $\ehi< \eth$ is defined to be underdense, for $\eth=0$,
 the underdense filling factor is zero, so is $N_C$. But when $\eth$ is raised above zero arbitrarily, a significant region with $\ehi=0=\rhi$ is identified as underdense, and $N_C$ as well as $\ffc$ of the underdense segment increases abruptly, as shown in figures \ref{fig:NC_vs_rhoth_void} and \ref{fig:FF_vs_rhoth} respectively. This discontinuous rise in underdense filling factor is higher for smaller redshift because the completely ionized segment ($\ehi=0$) grows with reionization\footnote{Although, for this threshold ($\ehi \gtrsim 0$), $FF_{\rm underdense}$ is significantly higher at $z=7$ than that at $z=10$, the completely ionized regions are more interconnected at $z=7$. Consequently, there are less number of underdense clusters at $z=7$ than at $z=10$.}.
 As we keep increasing $\eth$, the underdense segment at any redshift grows causing rise in both $N_C$ and $\ffc$. Again in the vicinity of percolation (in the underdense segment) $N_C$ starts to decrease while $\ffc$ keeps increasing as $\eth$ is increased. The percolation transition for underdense section at various redshifts is again shown by the vertical dotted lines in figure \ref{fig:NC_vs_rhoth_void} and by the filled circles in figure \ref{fig:FF_vs_rhoth}. Note that for $z \lesssim 9$, the completely ionized regions percolate, as described in \cite{Bag:2018zon}. Hence, the critical filling factor at the percolation transition in HI underdense segment cannot be defined at $z=7$. Instead, we show the percolation transition in underdense segment at $z=7$ by the red dotted vertical line in figure \ref{fig:FF_vs_rhoth}.
 
 In summary, for very low values of density threshold, the largest HI overdense cluster becomes percolating at all the redshifts, i.e., there exists a huge connected overdense cluster which is formally infinite in size. Similarly, above another critical threshold, the largest HI underdense cluster is percolating. The topology of either (HI overdense/underdense) segment before the respective percolation is of clumpy/bubble type, whereas after the percolation, the topology becomes network or sponge-like.
 There exists a range of $\eth$ for which both the HI overdense and underdense segments are percolating and the largest clusters in both the segments are formally infinite in extent.       

\end{itemize}

\subsection{Critical density threshold at percolation as a function of neutral fraction}

\begin{table*}
\begin{tabular}{|c|c|c|c|c|c|c|c|}
 \hline
 \multicolumn{1}{|c|}{\multirow{2}{*}{Redshift}} & \multicolumn{1}{|c|}{\multirow{2}{*}{$x_{\rm HI}$}}  & \multicolumn{6}{|c|}{Critical thresholds at percolation transition}\\ \cline{3-8}\rule{0pt}{1.5ex}
  & & \multicolumn{3}{|c|}{HI overdense} & \multicolumn{3}{|c|}{HI underdense} \\ \cline{3-8} 
  
   &  &\rule{0pt}{2.5ex}  $\ehic $ & $\dehic$ & $FF_{\rm overdense}^C$ & $\ehic$  & $\dehic$ & $FF_{\rm underdense}^C$ \\
 \hline
 \hline
 $7.0$ & $0.1484$ & $0.8891$ & $0.7407$ & $0.0407$ & $0$ & $-0.1484$  &-- \\
 \hline
 $7.5$ & $ 0.3243$& $0.9859$ & $0.6616$ & $0.0749$ & $0$ & $-0.3243$ &-- \\
 \hline
 $8.0$ & $ 0.4948$ & $1.0703$ & $0.5755$ & $0.1000$ & $0$ & $-0.4948$ &-- \\
 \hline
 $9.0$ & $0.7256$ & $1.2328$ & $0.5072$ & $0.1108$ & $0$ & $-0.7256$ &-- \\
 \hline
 $10.0$ & $0.8595$ & $1.3781$ & $0.5186$ & $0.0988$ & $0.5641$ & $-0.2954$ &$0.1854$ \\
 \hline
 $11.0$ & $0.9302$ & $1.5031$ & $0.5729$ & $0.0793$ & $0.5938$ & $-0.3364$ &$0.1482$ \\
 \hline
 $13.0$ & $0.9853$ & $1.5859$ & $0.6006$ & $0.0632$ & $0.6422$ & $-0.3432$ &$0.1265$ \\
 
 \hline
\end{tabular}
\caption{The table lists neutral fraction ($x_{\rm HI}$) and the critical threshold at the onset of percolation transition (for both overdense and underdense segments) for all the seven redshifts we studied. The overdense region  percolates above the listed critical density threshold  $\ehic$ and filling factor $\ffc^C$ while the underdense region percolates below the listed $\ehic$ which corresponds to the given $\ffc^C$. For $z \lesssim 9$, the completely ionized regions percolate (see \protect\cite{Bag:2018zon}), and during the percolation transition $\ffc_{\rm underdense}$ undergoes a discontinuous change, as explained earlier. Therefore, $\ffc^C$ for the HI underdense segment is not defined at redshifts $z \lesssim 9$.}
\label{table:percolation}
\end{table*}

\begin{figure*}
\centering
\subfigure[]{\includegraphics[width=0.47\textwidth]{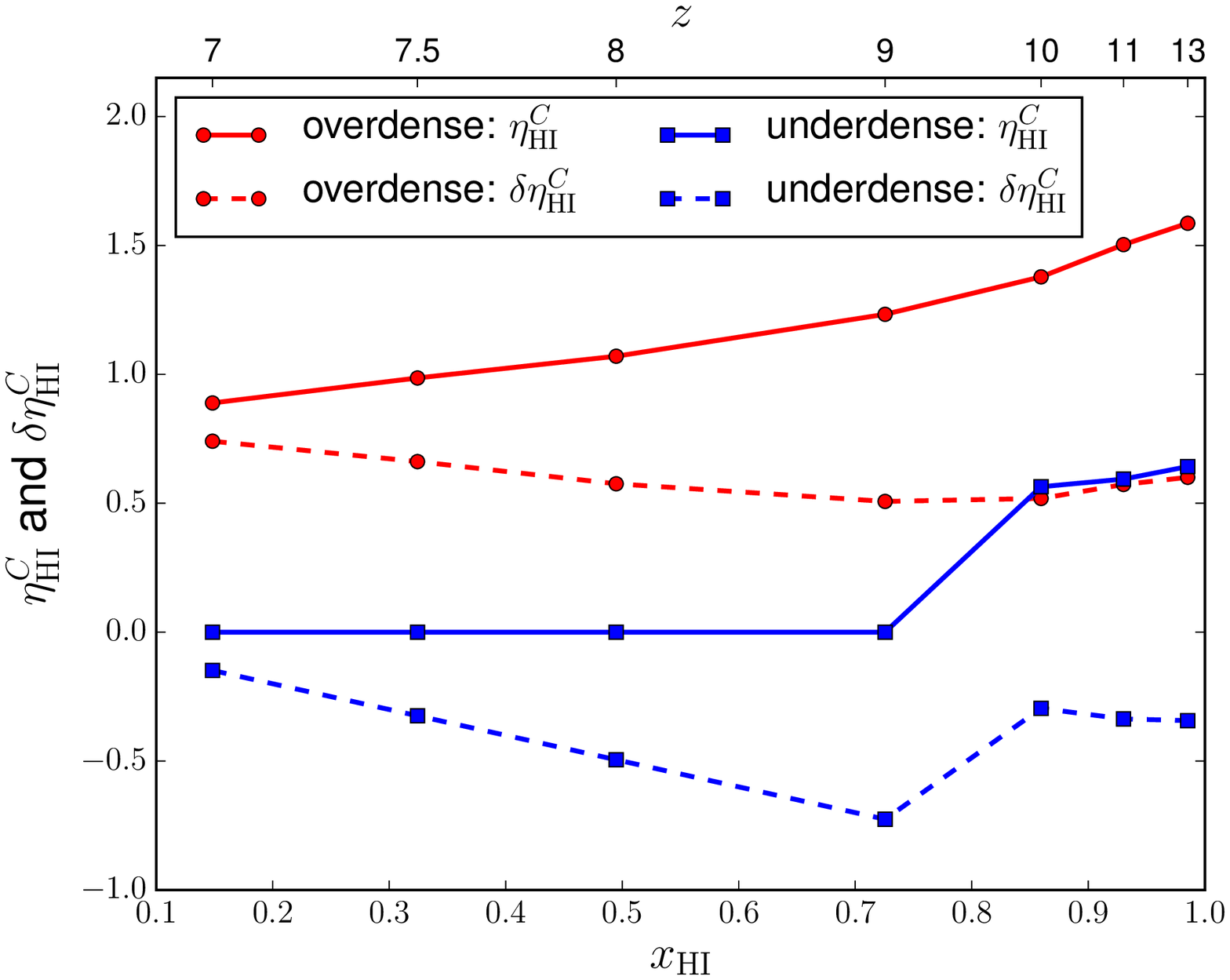}\label{fig:percolation_delta_xhi}}
\subfigure[]{\includegraphics[width=0.47\textwidth]{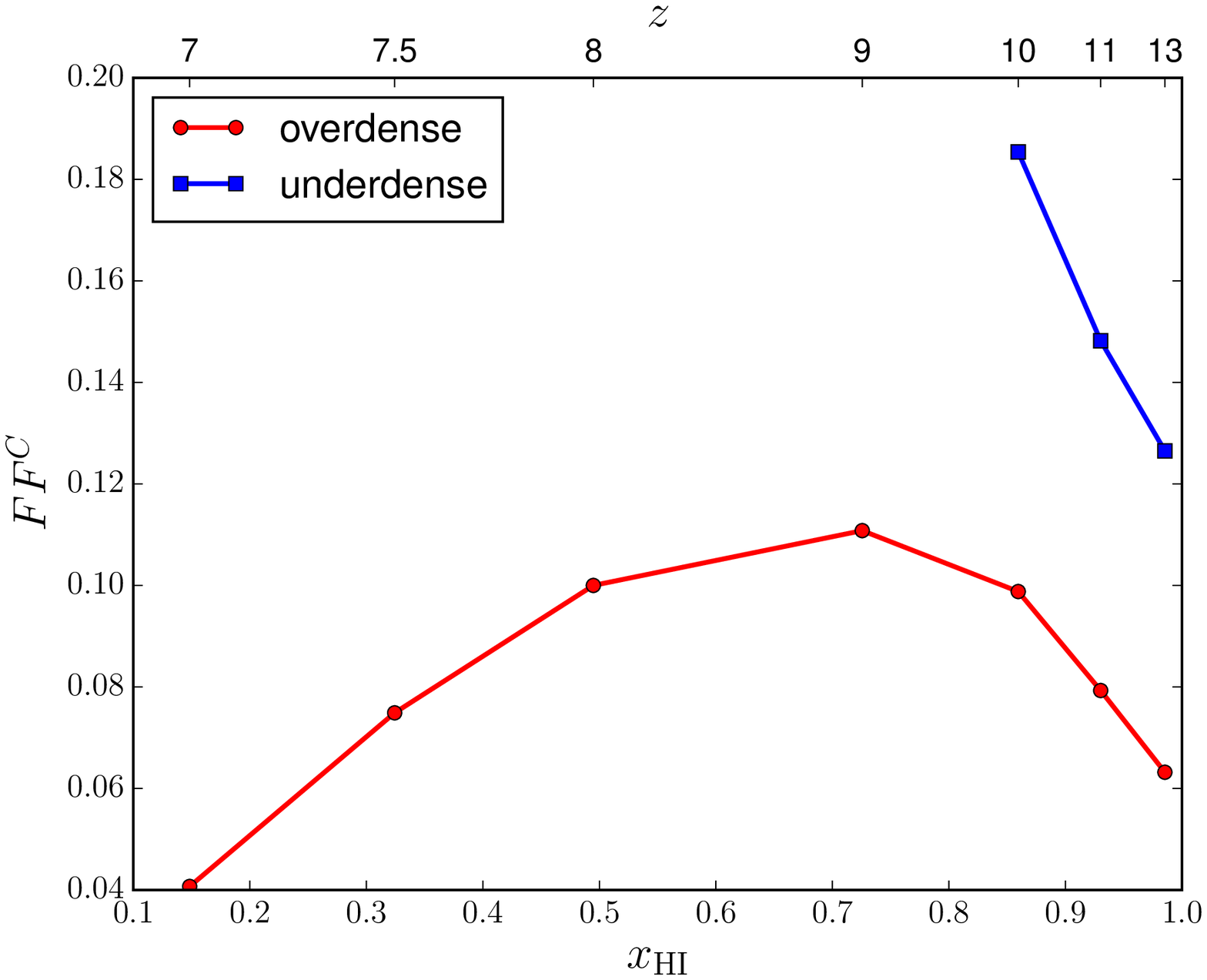}\label{fig:ffc}}
\caption{ {\bf (a):} The critical values of thresholds ($\ehic$ and $\dehic$) at the onset of percolation transition are plotted against corresponding $x_{\rm HI}$ for both overdense and underdense regions. The quantity $\dehi$ can be directly linked to the future observations in Fourier space. {\bf (b):} The critical filling factor $FF^C$ at the percolation transition for the HI overdense and underdense segments are plotted against $\xhi$. For underdense segment, $FF^C$ is not defined below $z\lesssim 9$ for which the ionized regions ($\ehi=0$) percolate. Interestingly the underdense segment percolates at a higher value of $FF^C$ compared to that of overdense segment. $FF^C_{\rm underdense}$ increases as reionization proceeds while $FF^C_{\rm overdense}$ exhibits a maximum.  Note that with the progress of reionization the neutral fraction $x_{\rm HI}$ decreases. The seven points in both the panels represent the seven different redshifts ($z=7,7.5,8,9,10,11,13$) shown along the top x-axis.}
\label{fig:dthC_vs_xHI}
\end{figure*}

The critical values of $\ehi$, $\dehi$ and $\ffc$ at the onset of percolation are listed in table \ref{table:percolation} (for both HI overdense and underdense segments) at various redshifts. In figure \ref{fig:percolation_delta_xhi}, the critical thresholds ($\ehic$ and $\dehic$) in both HI overdense and underdense segments are plotted against the neutral fraction $\xhi$ for the seven redshifts listed in table \ref{table:percolation}. The largest cluster in HI overdense segment is percolating for any threshold $\eth <\ehic$, equivalently for $\deth <\dehic$. On the other hand, the HI underdense segment percolates for $\eth >\ehic$, i.e. for $\deth <\dehic$.
At all the redshifts, the percolation transition in HI overdense segment takes place at a higher value of the threshold than that in the underdense segment. Interestingly, $\dehic$ for the overdense segment is always positive whereas we observe a negative value of $\dehic$ for the underdense segment at all stages of reionization. As redshift decreases, $\ehic$ for both HI overdense and underdense segments also decreases ($\dehic$ does not vary monotonically with $\xhi$).

Figure \ref{fig:ffc} shows plots of critical filling factor $FF^C$ at percolation in both HI overdense and underdense segments as a function of $\xhi$.  Interestingly, the $FF^C$ for the overdense segment exhibits a maximum at $\xhi \approx 0.7$. The maximum appears to be a feature of this type of reionization model and allows the predictions of the model to be tested against observations. $\ffc^C$ for the underdense segment increases with decreasing $\xhi$. Note that $\ffc_{\rm underdense}^C$ is not defined at $z \lesssim 9$, as explained in the previous subsection.   

For a Gaussian random field, the overdense and underdense segments are statistically identical. But in our case, the overdense segment percolates at lower values of filling factor (i.e. more easily) than the underdense segment at all redshifts. This observation is consistent with the results from similar analyses carried out on large-scale matter distribution in the universe and manifests the non-Gaussianity in the system originating from nonlinear gravitational clustering \citep{Yess1996, Sahni:1996mb}.

\subsection{Largest cluster statistics} \label{subsec:lcs}

\begin{figure*}
\centering
\subfigure[HI overdense segment]{
\includegraphics[width=0.47\textwidth]{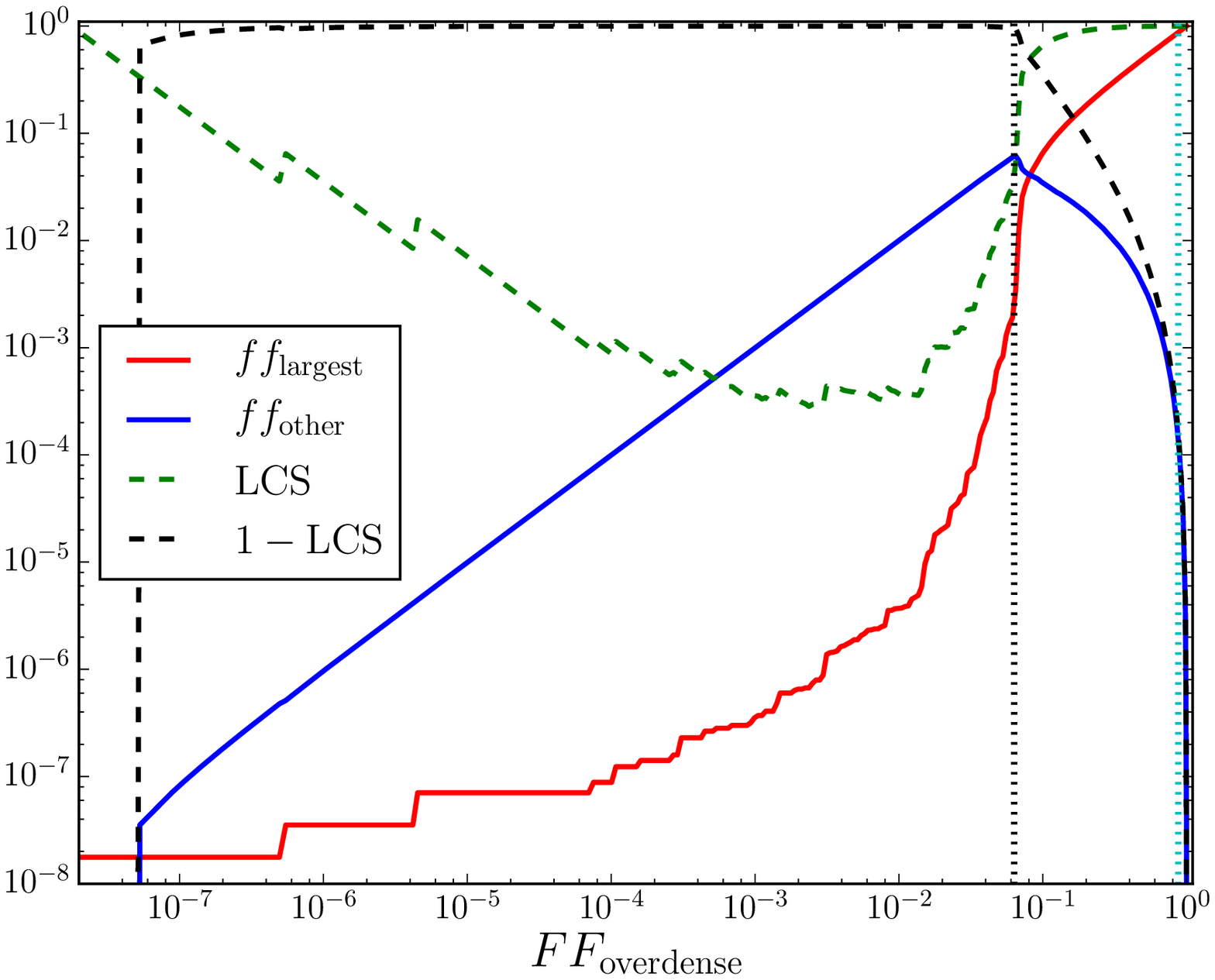}\label{fig:z13_HI_lcs}}
\subfigure[HI underdense segment]{
\includegraphics[width=0.47\textwidth]{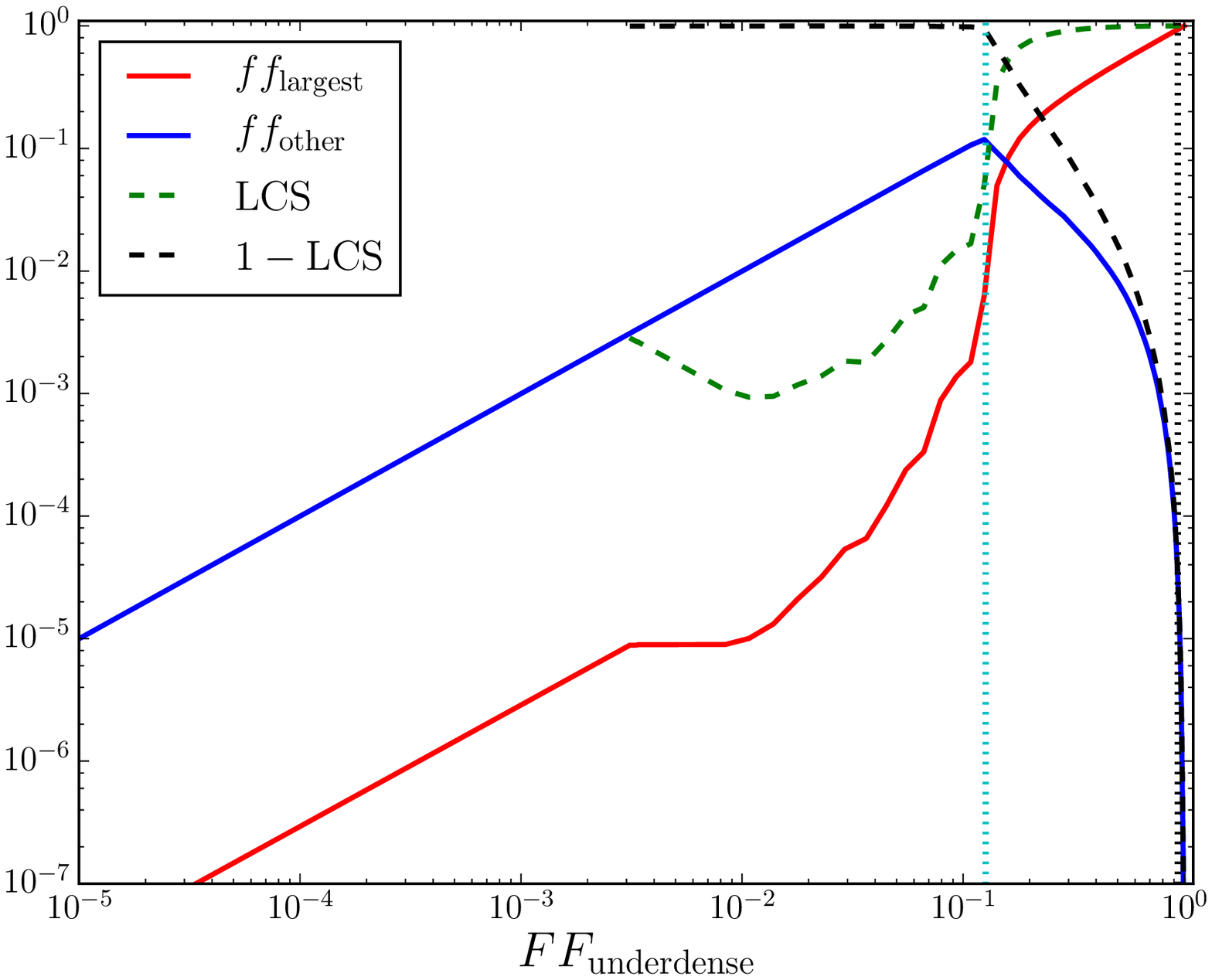}\label{fig:z13_ion_lcs}}
\caption{ The cluster statistics at $z=13$: The four fractions, $ff_{\rm largest}$, $ff_{\rm other}$, LCS and $(1-{\rm LCS})$ for HI overdense and underdense segments are plotted against corresponding filling factors in left and right panel respectively. In both panels the percolation transition for overdense region is shown by black dotted vertical line  while the cyan dotted vertical line represents the percolation in the underdense segment. In both the cases, during percolation, the fractions $ff_{\rm largest}$ and LCS rise steeply while the other two fractions $ff_{\rm other}$ and   $(1-{\rm LCS})$ decay sharply. These sharp changes in these `ordered parameters' can also independently define the percolation transition in either segment, as explained by \protect\cite{Klypin1993, Yess1996}.}
\label{fig:lcs_z13}
\end{figure*}

We now study the percolation process using the  ``largest cluster statistics'' (LCS) which is the fraction of the volume (HI overdense or underdense) filled by the largest cluster \citep{Klypin1993}. In this analysis, a key role is played by the following quantities which are defined, for a given density threshold, for both the HI overdense and underdense excursion sets separately:
\begin{equation}\label{eq:lcs_HI}
  {\rm LCS}=\frac{\rm volume ~of ~the ~largest~cluster}{\rm total ~volume ~of ~all ~the ~clusters}\;,
\end{equation}
\begin{align}\label{eq:fp_largest}
  ff_{\rm largest} &=\frac{\rm volume ~of ~the ~largest ~cluster}{\rm volume ~of ~the ~simulation ~box}=\ffc \times {\rm LCS}\;,
\end{align}
\begin{align}
 ff_{\rm other}& =\frac{\rm volume ~of ~all ~clusters ~other ~than ~the ~largest ~cluster}{\rm volume ~of ~the ~simulation ~box} \\ 
 &\equiv \ffc -ff_{\rm largest}=\ffc \times (1-{\rm LCS}) \;. \nonumber
\end{align}
The fractions $ff_{\rm largest}$ and $ff_{\rm other}$ are essentially the filling factors of the largest cluster and the rest of the clusters (all the clusters excluding the largest cluster) respectively.

\begin{figure*}
\centering
\hspace{-4.0 mm}
\subfigure[$z=13$]{
\includegraphics[width=0.35\textwidth]{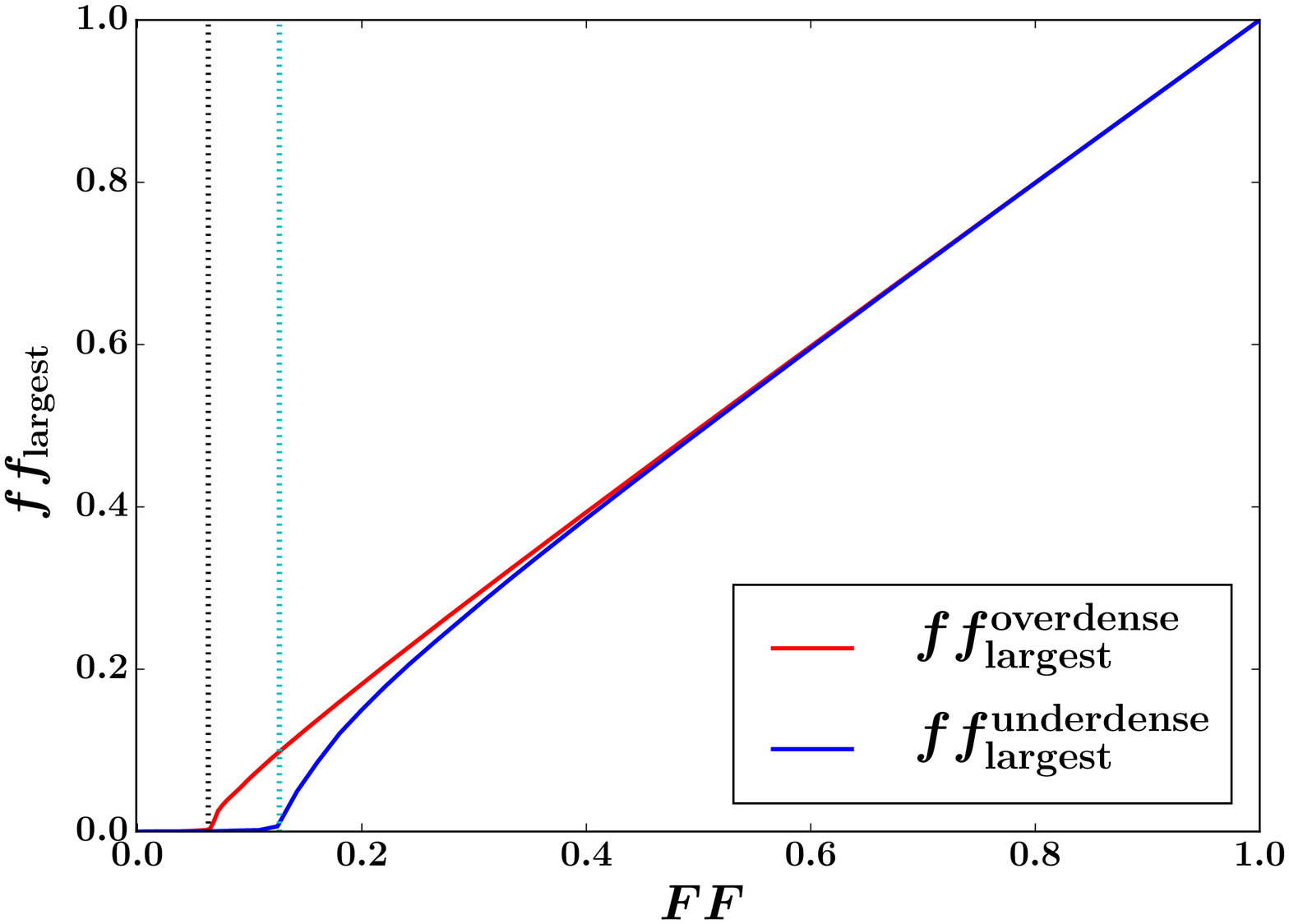}}\hspace{-7 mm}
\subfigure[$z=11$]{
\includegraphics[width=0.35\textwidth]{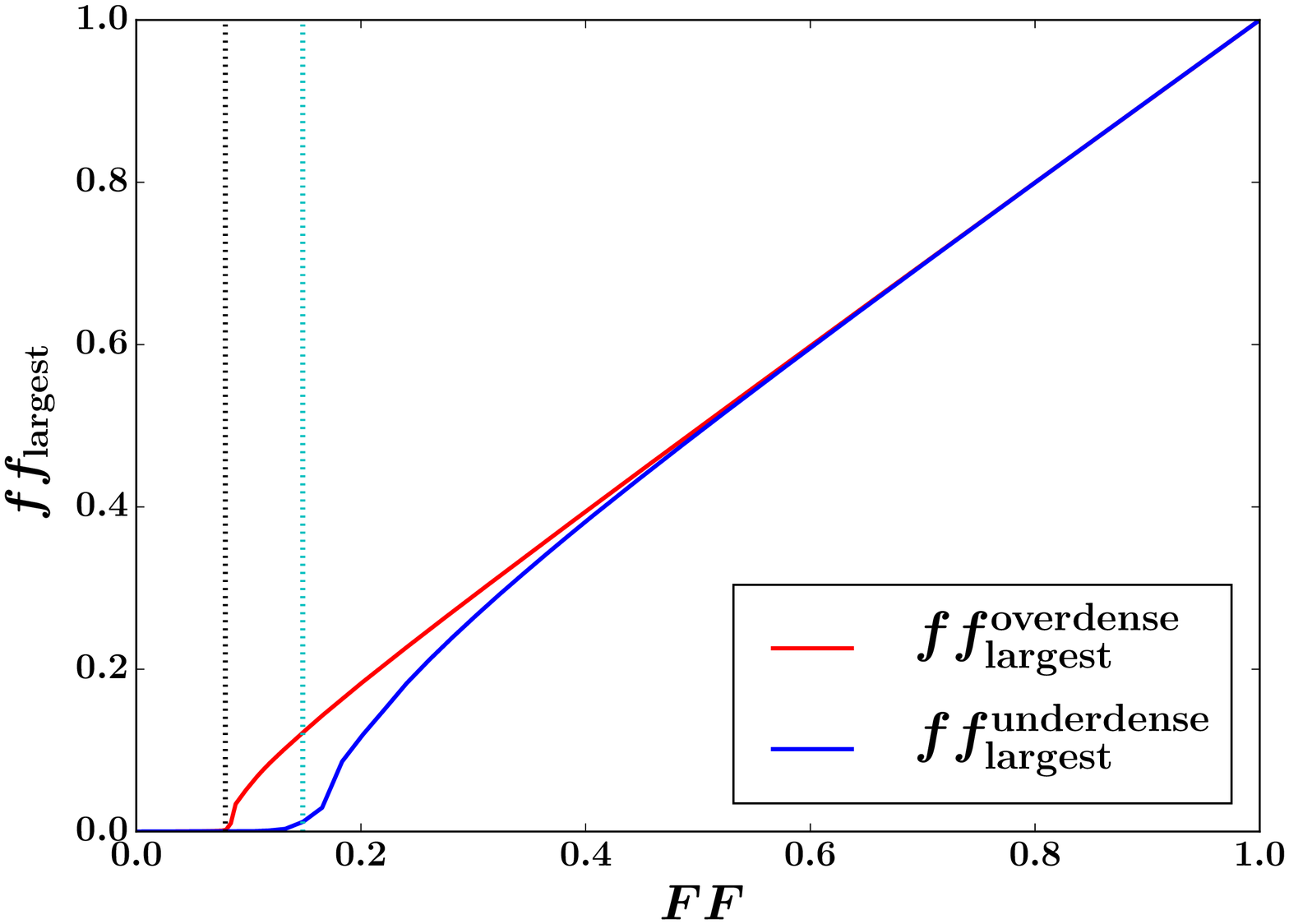}}\hspace{-7 mm}
\subfigure[$z=10$]{
\includegraphics[width=0.35\textwidth]{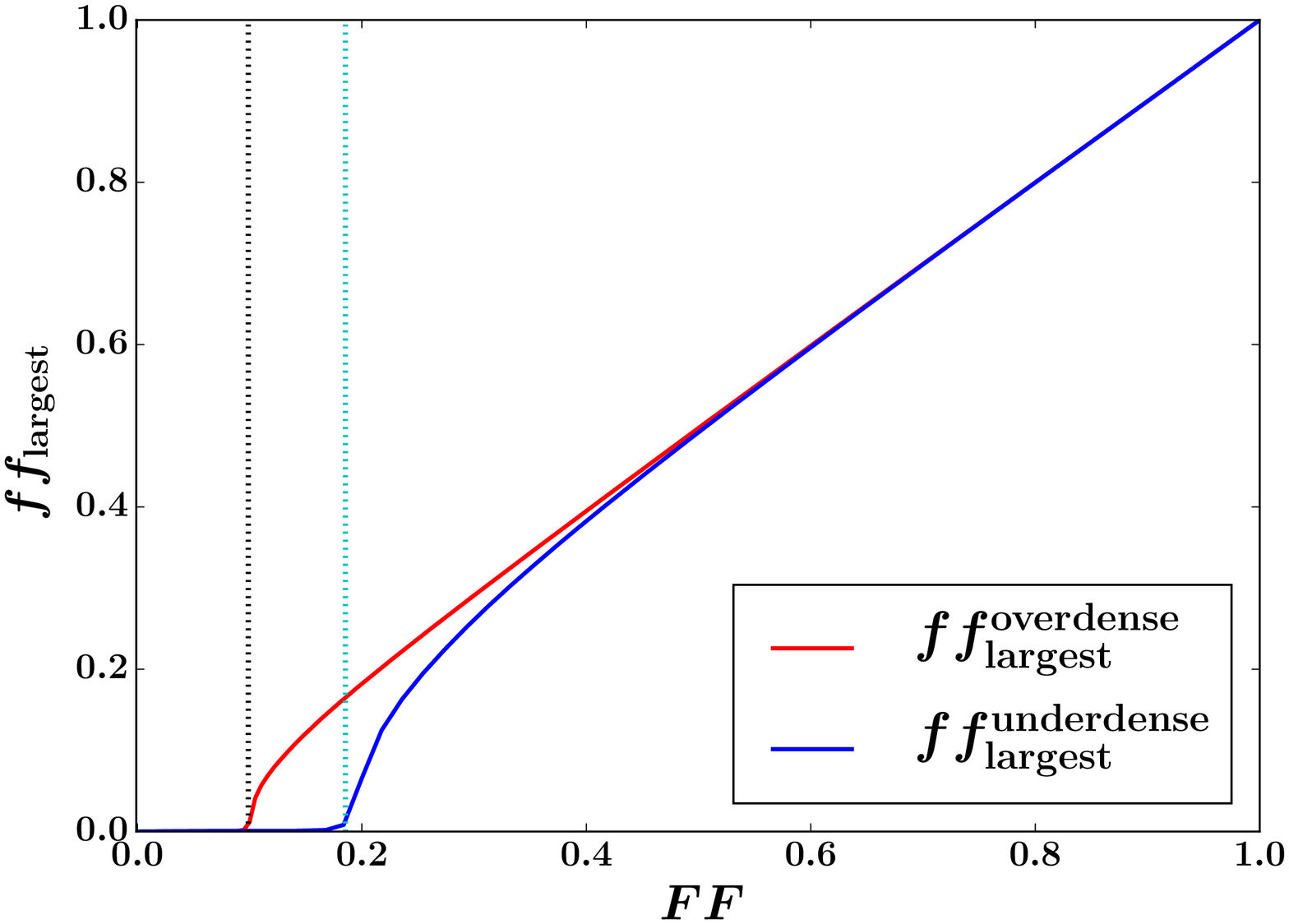}}\hspace{-4.0 mm}
\caption{ Percolation curves at $z=13,11,10$. For Gaussian random fields (GRF) overdense and underdense segments percolate identically at same filling factor, ( e.g. at $\ffc \approx 0.16$ for a 3D GRF). But here the overdense and underdense components percolate at different values of filling factor. This asymmetry in the percolation processes in HI overdense and underdense segments or equivalently the area under the ``hysteresis'' in the percolation curves represents the amount of non-Gaussianity \protect\citep{Sahni:1996mb}. As we decrease the redshift from $z=13$, the area under the hysteresis increases indicating that the amount of non-Gaussianity is grows at later times, as expected, even in the HI field.}
\label{fig:pc}
\end{figure*}

For percolation analysis, in figure \ref{fig:lcs_z13}, the four fractions, $ff_{\rm largest}$, $ff_{\rm other}$, LCS and $(1-{\rm LCS})$, of both the HI overdense region (left panel) and the underdense region (right panel) are plotted against the corresponding filling factor at $z=13$. The percolation transition in the overdense/underdense segment is shown by the back/cyan vertical dotted line in both panels. The panels show that LCS  and $ff_{\rm largest}$ for both overdense and underdense segments rise sharply during the percolation transition while $(1-{\rm LCS})$ and $ff_{\rm other}$ sharply decay. This steep rise in the `ordered parameters', LCS and $ff_{\rm largest}$ (or the sharp decrease in $(1-{\rm LCS})$ and $ff_{\rm other}$), provides us an independent definition of the percolation transition in either segment, as explained by \cite{Klypin1993, Yess1996} \footnote{The fact that LCS rises sharply to order of unity during percolation transition implies that most of the overdense/underdense volume is suddenly occupied by the largest cluster beyond the percolation in the respective segment.}.

As discussed earlier, it can be noticed in figures \ref{fig:ffc}, \ref{fig:lcs_z13} that the overdense and underdense segments percolate at different values of respective filling factors which are also explicitly listed in table \ref{table:percolation}. It is well known that for Gaussian random fields (GRF), overdense and underdense segments percolate symmetrically at the same value of filling factor. For example, for a smooth GRF in 3D, the filling factors at percolation is $\ffc^C \approx 16 \%$ (but this value is expected to be affected by the finite size of the box and the discreteness of the grid).  
Hence this difference in the critical filling factors for HI overdense and underdense segments is a direct consequence of the non-Gaussianity in the HI density fields. The departure from the GRF can be more efficiently demonstrated by the `{\em percolation curves}' \citep{Sahni:1996mb}, which are  the plots of filling factor of the largest cluster ($ff_{\rm largest}$ defined in \eqref{eq:fp_largest}) against the respective total filling factor ($\ffc$) for both HI overdense and underdense excursion sets. 


To explore this further, we plot the percolation curves at different redshifts\footnote{Since for $z\lesssim 9$, the underdense $\ffc$ at percolation transition is undefined, hence we do not show the percolation curves in these range of redshift.}, $z=13,11~{\rm and}~ 10$ in figure \ref{fig:pc}. The figure shows that the area under the `hysteresis' actually increases with decreasing redshift indicating the significant increase of non-Gaussianity in the HI field as the universe evolves through reionization as well as structure formation in time. In recent times, efforts have been made to analyze the amount of non-Gaussianity in the 21-cm fields using higher order correlation functions \citep{Bharadwaj2005, Cooray2005, Pillepich2007, mondal2015, Yoshiura:2014ria, majumdar2017}. Complementary to the bispectrum statistics, in this paper, the growth of non-gaussianity with reionization has been demonstrated from the geometrical point of view using the percolation curves.

\section{Determining Shapefinders at varying values of \texorpdfstring{$\eth$}{Lg}}\label{sec:shape}

In this section we study the shape and morphology of the clusters in HI overdense and underdense segments using Shapefinders from excursion set approach for different redshifts listed in table \ref{table:percolation}. In principle, we could calculate the Shapefinders of all the clusters individually by triangulating their surfaces. However the surfaces of extremely small clusters, having only a few grid points inside them, could not be accurately modelled by the triangulation scheme. These small clusters are less interesting because they are expected to be mostly spherical and their Shapefinders calculated using any surface modelling scheme are not accurate enough due to the coarse resolution. Moreover, since the number of clusters is very high near the percolation transition (see figure \ref{fig:NC_rhoth_FF}), calculating Shapefinders for all the clusters will require enormous computation resources. Hence we consider only sufficiently large clusters, consisting of at least $100$ grid points, for triangulation. Since the definitions of the overdense and underdense segments in the HI density field, given by equation \eqref{eq:segments}, are subject to a density threshold $\eth$, the shape distribution of clusters in either segment crucially depends on the choice of $\eth$. One density threshold of interest is the critical threshold, just before percolation (for both HI overdense and underdense segments), as suggested by \cite{Shandarin:2003tx}, because the number of large clusters becomes maximum near that threshold. Indeed, well before percolation, the clusters (overdense or underdense) are quite small in size and also less in number. On the other hand, well after percolation there are only few sufficiently large clusters apart from the huge percolating cluster. Moreover, the computed Shapefinders for the percolating cluster suffer from inaccuracies because the cluster is not physically bounded, as explained in figure 1 of \cite{Bag:2018zon}. Hence the statistical study of the Shapefinders of clusters (both in HI overdense and underdense segments) is interesting in the vicinity of percolation. Therefore, we mainly focus on the threshold corresponding to the onset of percolation while studying the shape distribution of clusters in HI overdense/underdense excursion sets.

\subsection{HI overdense segment}\label{subsec:over}

\begin{figure*}
\centering
\subfigure[Thickness, breadth and length]{
\includegraphics[width=0.483\textwidth]{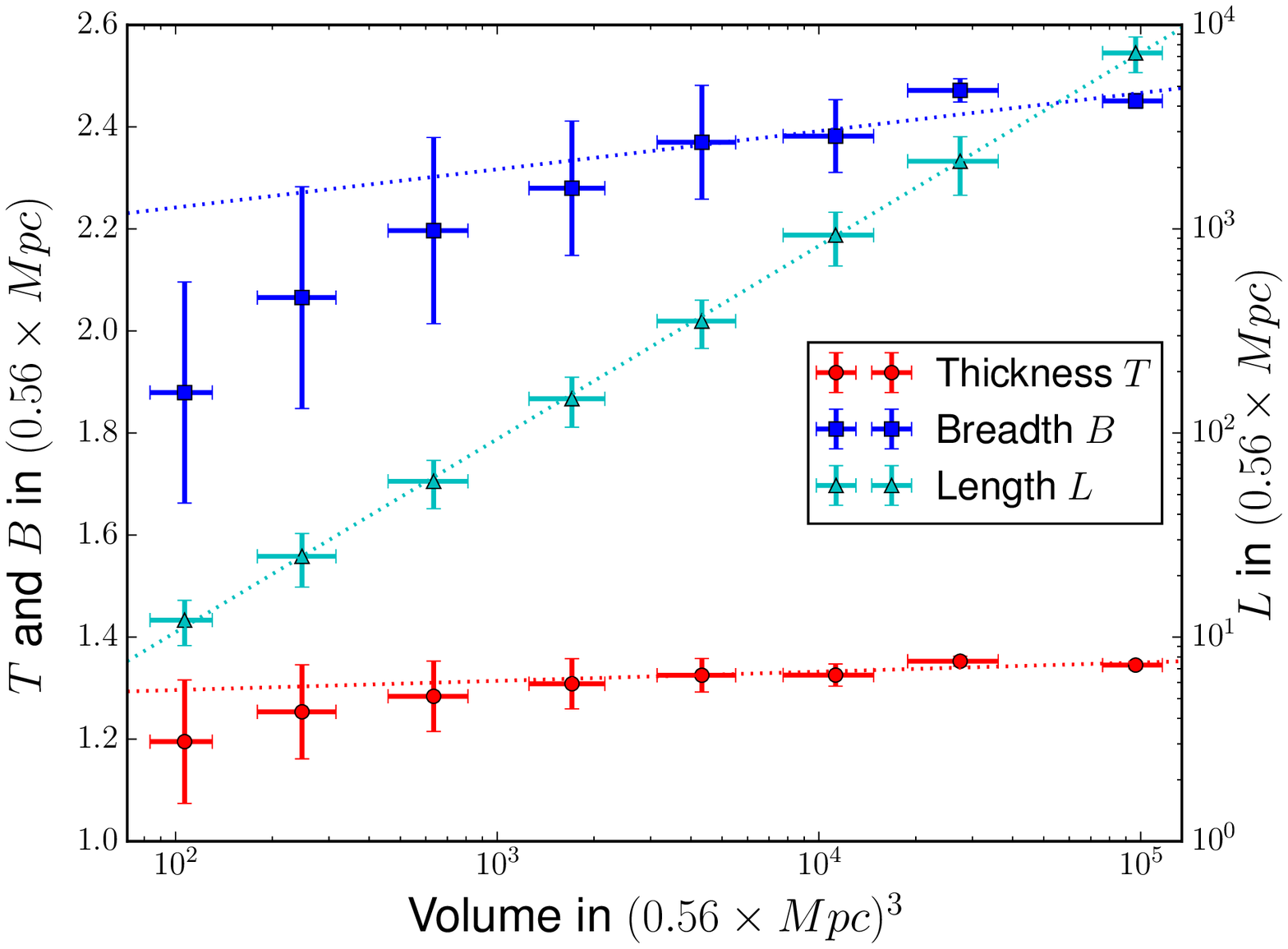}\label{fig:bin0_TBL_z13_iso101_5_c100}}
\subfigure[Planarity and filamentarity]{
\includegraphics[width=0.483\textwidth]{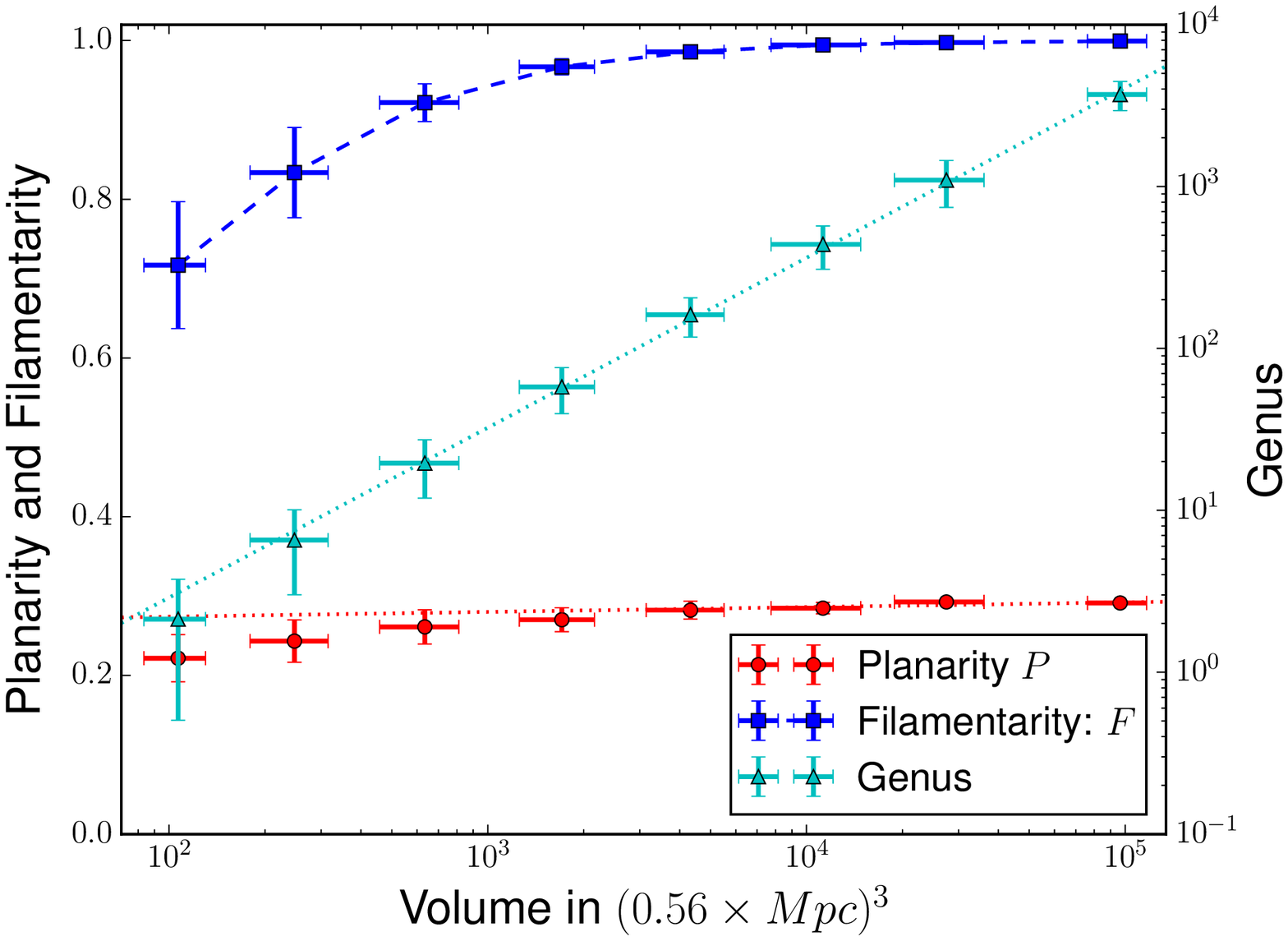}\label{fig:bin0_PF_z13_iso101_5_c100}}
\caption{ Shapefinders, planarity, filamentarity and genus of HI overdense clusters at $z=13$ are binned in volume for the threshold $\eth=1.5859$ which corresponds to just before percolation. Note that the values of Shapefinders are quoted in comoving scale.  In the left panel $T$, $B$ and $L$ are fitted to straight lines shown by the dotted lines with respective colours. The slopes of the best fit straight lines to $T$, $B$ are negligible. On the other hand, the best fit straight line to $\log L$ vs $\log V$ has slope $0.945 \pm 0.01$ which is $\mathcal{O}(1)$. This shows $T$ and $B$ of sufficiently large clusters increase very slowly with the cluster volume $V$ but $L$ almost increases linearly $V$.  From the right panel it is evident that the filamentarity increases with cluster volume and almost saturates to unity for large clusters. Filamentarity in volume bins are joined by dashed line for visual guidance. But the planarity does not increase much, hence the best fit straight line again has a negligible slope. Genus of clusters also increases almost linearly with the volume.}
\label{fig:struc_PF_z13}
\end{figure*} 

Plotting the Shapefinders of each individual cluster will make the figures unnecessarily busy and complicated. Hence, to study the Shapefinders statistically,  we bin the data (with bins equispaced in log scale) and plot the volume averaged values of Shapefinders, planarity, filamentarity and genus in the volume bins, as shown in rest of the figures in this section. 
The values of Shapefinders are quoted in comoving scale through out the paper.
 The standard deviations (weighted with volume of clusters) in each bin are shown as respective error bars which provide us with some measure of the scatter in each bin.

In figure \ref{fig:struc_PF_z13}, the Shapefinders, planarity, filamentarity and genus of HI overdense clusters are plotted in volume bins at the onset of percolation at $z=13$. The figure \ref{fig:bin0_TBL_z13_iso101_5_c100} shows that larger clusters have similar values of thickness ($T$), as well as breadth ($B$). But the length ($L$) of the clusters, plotted in log scale along the right y-axis, increases almost linearly with the volume ($V$) of the clusters since $V \propto TBL$. Note the enormous difference in scales of $T$ and $B$ on one hand and $L$ on the other. The best fit straight lines are shown by the dotted lines with respective colours\footnote{The dotted lines represent the best fit straight lines whereas the dashed lines join the data points for visual guidance in the plots of Shapefinders, planarity, filamentarity and genus, throughout the paper.}. The slopes of the best fit straight lines to $T$ and $B$ are negligible while the slope of the best fit straight line to  $ \log L$ vs $\log V$ curve is of order unity as expected\footnote{For spherical surfaces the slope of $ \log L$ vs $\log V$ line would be $m_L=1/3$, for sheets $m_l = 1/2$ while for filaments $m_L=1$.}, namely $m_L \equiv  \dd \log L / \dd \log V = 0.945 \pm 0.01$. In the figure \ref{fig:bin0_PF_z13_iso101_5_c100}, the planarity ($P$), filamentarity ($F$) and genus ($G$) of the clusters are plotted in volume bins. Since thickness and breadth of the large clusters do not vary much with volume, the planarity, defined in \eqref{eq:G}, is also similar for large clusters, as illustrated in the figure \ref{fig:bin0_PF_z13_iso101_5_c100}. Moreover the thickness of a large cluster is of same order of magnitude as its breadth which results in low value of the planarity of the large cluster. On the other hand, the filamentarity of the clusters increases with volume and saturates to almost unity for large clusters. Therefore, statistically the large clusters are highly filamentary.  This is due to the fact that for the larger clusters $L \gg T,B$. The genus, plotted in log scale along right y-axis in figure \ref{fig:bin0_PF_z13_iso101_5_c100}, also increases with cluster size indicating that the clusters become more porous with increasing size as more tunnels pass through them. In this plot the filamentarity is not fitted with straight line, instead the data points are joined by the dashed lines for visual guidance.

\begin{figure*}
\centering
\subfigure[$z=8$, $\eth=1.0703$]{
\includegraphics[width=0.504\textwidth]{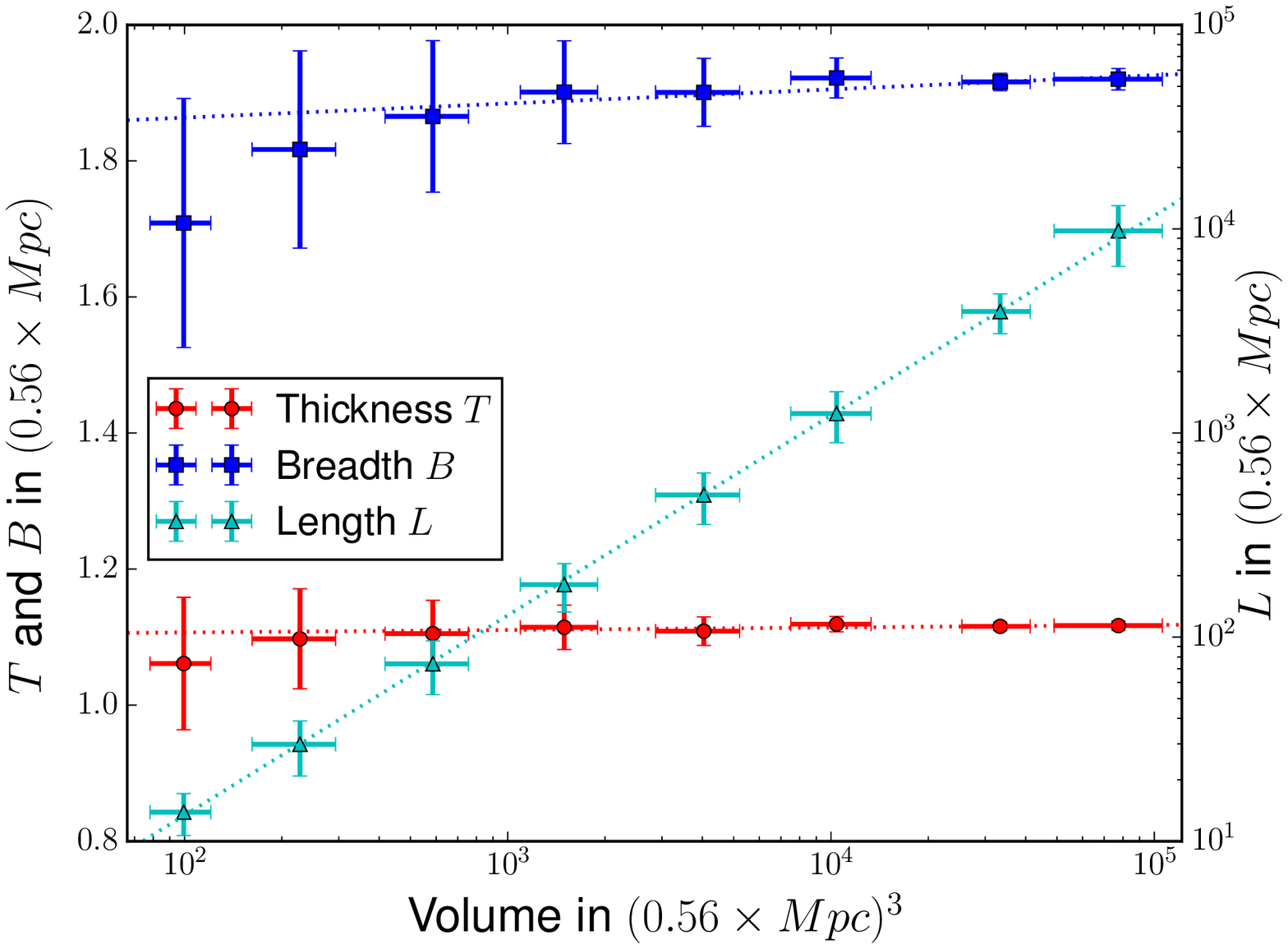}\label{fig:bin0_TBL_z8_iso68_5_c100}}
\subfigure[Distribution of Cross-section]{
\includegraphics[width=0.464\textwidth]{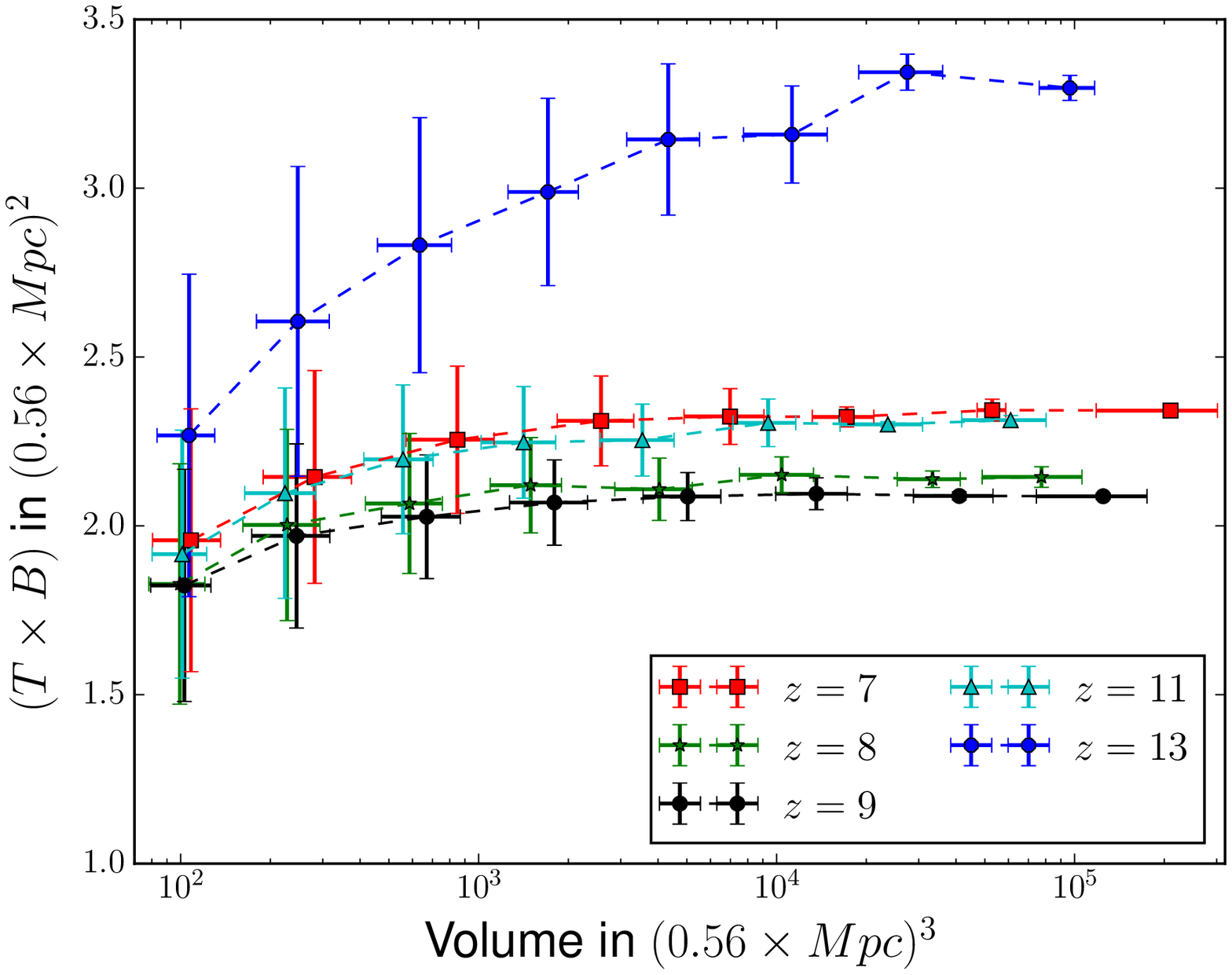}\label{fig:Cross_TB}}
\caption{ {\bf (a):} In the left panel, Shapefinders $T$, $B$, $L$ of clusters in HI overdense region are plotted in volume bins for $z=8$ at the onset of percolation. {\bf (b):} In the right panel, we plot $T \times B$, which can be interpreted as the `cross-section' of a filamentary cluster, in volume bins just before percolation in the HI overdense segment at various redshifts. All the curves flattens at higher volume bins while the error bars shrink. Thus, it is evident that the large clusters at each redshift possess a characteristic cross-section. The similarity of cross-section among large HI overdense clusters become more pronounced at lower redshifts, i.e. at advanced stages of reionization.}
\label{fig:struc_TBL_z8_TB}
\end{figure*}

Focusing on later stages of reionization we plot the Shapefinders of HI overdense clusters at a lower redshift, $z=8$, in the left panel of figure \ref{fig:struc_TBL_z8_TB}. Again the choice of the density threshold corresponds to the onset of percolation in the HI overdense segment at $z=8$. A visual comparison between the figures \ref{fig:bin0_TBL_z8_iso68_5_c100} and \ref{fig:bin0_TBL_z13_iso101_5_c100} reveals that the thickness and specially breadth of large clusters at $z=8$ are more insensitive to volume when compared with that of $z=13$. Note that, the behaviour of the planarity, filamentarity and genus of HI overdense clusters with the cluster volume at $z=8$ is similar to that of $z=13$, shown in \ref{fig:bin0_PF_z13_iso101_5_c100}. The product of the first two Shapefinders, $T \times B$, can be interpreted as the ``cross-section'' of a filamentary cluster. In figure \ref{fig:Cross_TB}, the cross-sections of clusters are plotted in volume bins at different redshifts. For each redshift we set $\eth=\ehic$, i.e. we choose the density threshold corresponding to the onset of percolation in HI overdense segment. It is evident that the cross-section of large clusters are alike. The similarity of cross-section among large HI overdense clusters become more pronounced at lower redshifts, i.e. at advanced stages of reionization.

\begin{figure*}
\centering
\subfigure[]{
\includegraphics[width=0.483\textwidth]{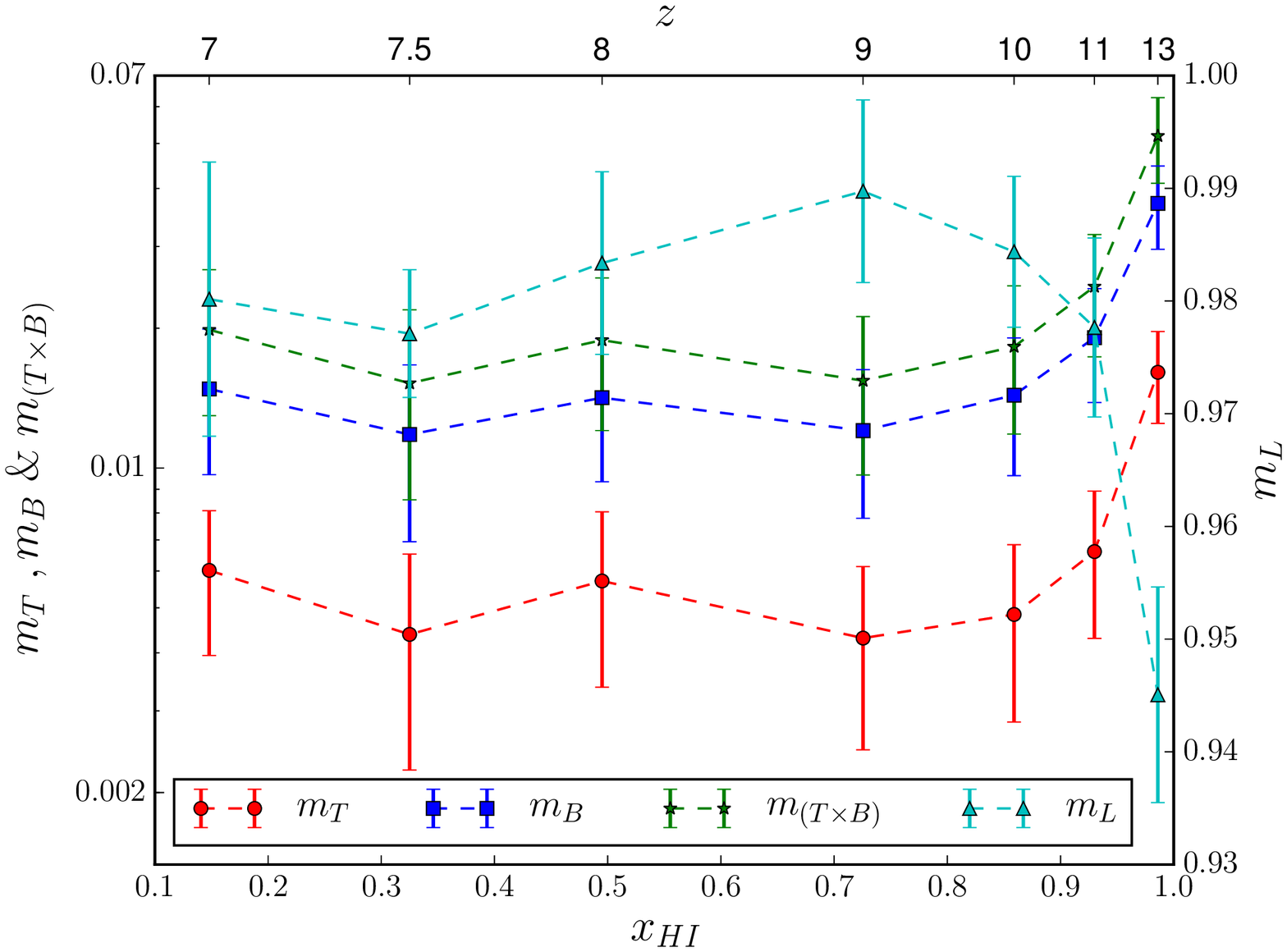}\label{fig:m_TBL}}
\subfigure[]{
\includegraphics[width=0.483\textwidth]{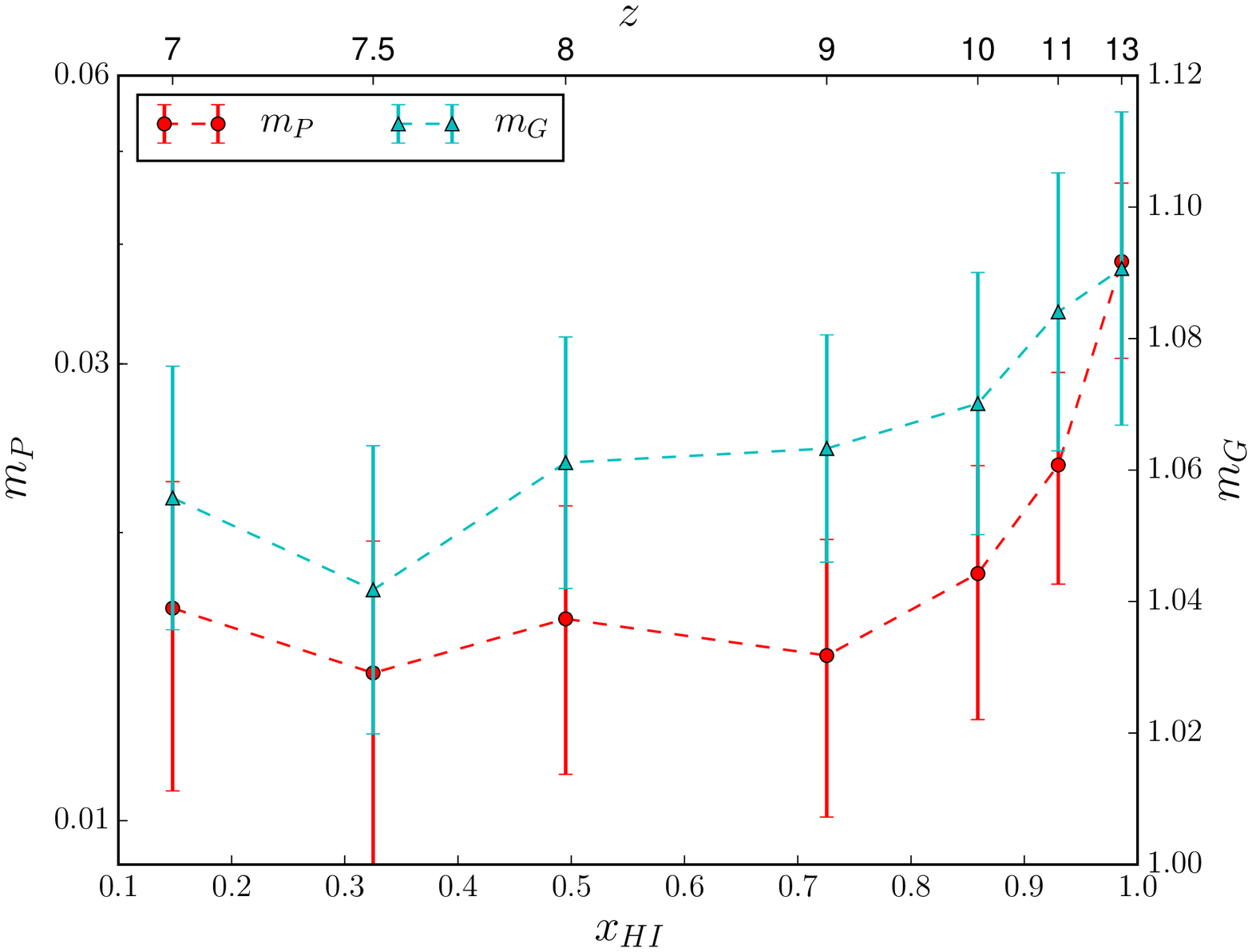}\label{fig:m_PG}}
\caption{ The slopes of the best fit straight lines to Shapefinders of HI overdense clusters are plotted for various neutral fractions. The slopes are defined as $m_X \equiv \dd \log X/\dd \log V$ where $X \in (T,B, T\times B, L,P,G)$. As reionization proceeds, $m_T$, $m_B$, $m_{(T \times B)}$ decreases significantly in log-scale. This implies that the thickness, breadth and cross-section of HI overdense clusters become more similar at advanced stages of reionization. On the other hand, $m_L$ increases with reionization and attains to almost unity at later stages of reionization when $L \propto V$. {\bf (a)} In the left panel, the slopes $m_T$, $m_B$, $m_{T \times B}$ are plotted along left y-axis in log scale while $m_L$ is plotted along right y-axis in linear scale.  {\bf (b)} In the right panel, $m_P$ is plotted in log scale along left y-axis whereas $m_G$ is plotted along right y-axis in linear scale. The decrease in $m_P$ with reionization is supported by the fact that at lower redshifts large clusters have more alike values of $T$, as well as $B$. $m_G$ is of order unity at all redshifts. This indicates that genus values of clusters are almost proportional to their lengths, $G \propto V \propto L$.}
\label{fig:slopes_overdense}
\end{figure*}

The slopes of the best fit straight lines to the Shapefinders, planarity and genus of HI overdense clusters have been plotted against neutral fraction in figure \ref{fig:slopes_overdense}. The redshifts corresponding to the neutral fractions are listed in table \ref{table:percolation} and shown along the top x-axis. The slopes are defined as $m_X \equiv \dd \log X/\dd \log V$ where $X$ belongs to the set $(T,B,T \times B,L,P,G)$. From figure \ref{fig:m_TBL} it is clearly evident that as reionization begins, $m_T$, $m_B$ and $m_{T \times B}$ decreases. This implies that the thickness, breadth and the cross-section of HI overdense clusters, just before percolation, becomes more insensitive to volume as reionization proceeds. Therefore, $m_L$ increases to almost unity at lower redshifts. Figure \ref{fig:m_PG} shows that $m_P$ also decreases with reionization. Interestingly, $m_G$ is of order unity which implies that the genus increases almost linearly with cluster volume at all redshifts, i.e. $G \propto V \propto L$.

In summary the following conclusions can be noted regarding Shapefinders of HI overdense clusters:
\begin{itemize}
\item  At the onset of percolation transition, many large clusters appear in the maps making it ideal for statistical study of Shapefinders. Therefore, we focus on just before percolation (i.e. setting $\eth \approx \ehic$) for each case.

\item The error bars on $T$, $B$, $P$, $F$ also shrink significantly in larger volume bins. This indicates that the Shapefinders of individual smaller clusters are more scattered, while the larger clusters have more similar values of the Shapefinders -- $T$ and $B$, as well as $P$ and $F$. 

 \item We observe that larger clusters are statistically more filament-like at all redshifts. It is also noticed that the large clusters possess similar values of thickness ($T$), as well as breadth ($B$). However, their third Shapefinder -- length ($L$) -- increases almost linearly with their volume, $L \propto V$. Moreover the thickness ($T$) of a large clusters is of same order as its breadth ($B$) which ensures that planarity ($P$) of the large cluster is quite small. Since $T$ and $B$ do not vary much with cluster size, $P$ also remains stable.

 \item Larger clusters tend to have similar cross-section (estimated by $T \times B$) and only their ``length'' varies linearly with volume. This is expected since, as we lower the density threshold, the large clusters grow effectively through merging of many relatively smaller filamentary clusters which were themselves large enough to have similar $T$ and $B$ (hence similar cross section). This justifies the overwhelming filamentarity of larger clusters.  Also it is evident that these filaments at various redshifts (and various threshold) have cross-section of same order of magnitude.

 \item The higher values of genus in larger clusters indicate that the large clusters have highly non-trivial topology. As we lower the density threshold, the large clusters grow as many filamentary branches and sub-branches connect to them. Thus the large clusters acquire multiply connected structure with many tunnels (of HI underdense regions) passing through them.     
 
 
 \item The similarity in $T$ or $B$ among large clusters, just before percolation, becomes more profound at lower redshifts. Therefore, as reionization proceeds the cross-sections of large clusters become more alike.
 
\end{itemize}


\subsection{Underdense regions in HI}\label{subsec:under}

\begin{figure*}
\centering
\subfigure[Thickness, breadth and length]{
\includegraphics[width=0.483\textwidth]{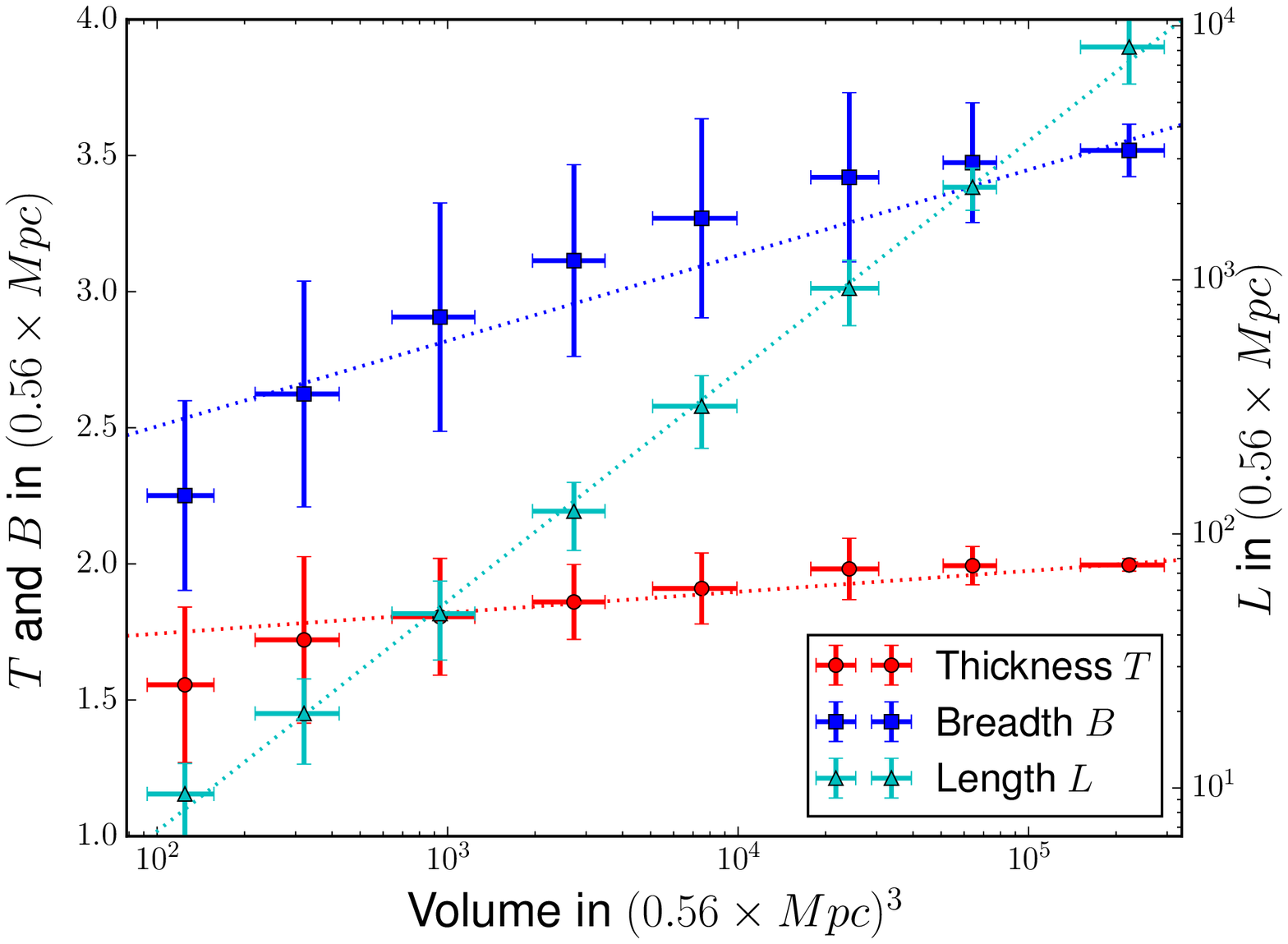}\label{fig:void_bin0_TBL_z13_iso41_1_c100}}
\subfigure[Planarity and filamentarity]{
\includegraphics[width=0.483\textwidth]{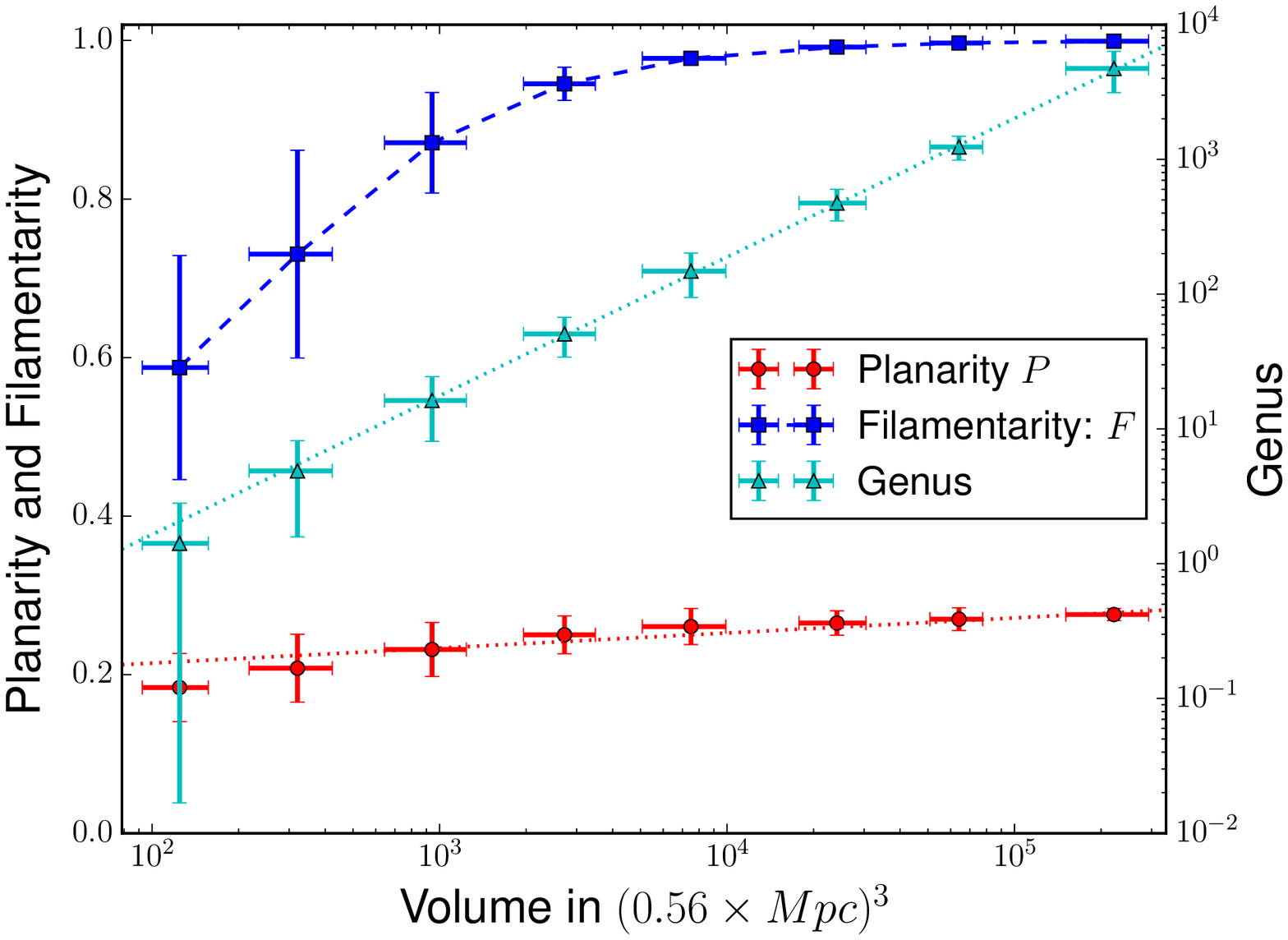}\label{fig:void_bin0_PF_z13_iso41_1_c100}}
\caption[ The Shapefinders (left panel), planarity, filamentarity and genus (right panel) of the clusters in HI underdense region are plotted in volume bins for $z=13$ at the onset of percolation; i.e. for the density threshold being set at $\eth=0.6422$.]{ The Shapefinders (left panel), planarity, filamentarity and genus (right panel) of the clusters in HI underdense region are plotted in volume bins for $z=13$ at the onset of percolation; i.e. for $\eth=0.6422$.}
\label{fig:void_TBL_z13}
\end{figure*}

 To complement the morphology of the HI overdense segment, we also study the Shapefinders of clusters in the underdense segment\footnote{Note that the `clusters' defined here, being underdense, are formally local `voids' in the HI density fields.}, as defined by the density threshold $\eth$ in equation \eqref{eq:segments}. Again we focus on the onset of percolation in the underdense segment and calculate the Shapefinders of sufficiently large clusters (having at least 100 underdense grid points inside each). The Shapefinders, binned in volume, of HI underdense clusters at $z=13$ are plotted in figure \ref{fig:void_bin0_TBL_z13_iso41_1_c100}. Again we observe that thickness ($T$), as well as breadth ($B$), of underdense large clusters remain quite insensitive to the cluster volume but the length $L$ increases with volume. This leads to increase in filamentarity with cluster volume while the planarity does not vary much, as shown in \ref{fig:void_bin0_PF_z13_iso41_1_c100}. 
 
 We also study the Shapefinders of HI underdense clusters at lower redshifts. As explained in \cite{Bag:2018zon}, at $z\lesssim 9$, the underdense segment  percolates for any non-zero value of density threshold, i.e. for any $\eth>0$. Since the Shapefinders of the large percolating cluster may not be calculated precisely, the study of morphology of HI underdense clusters at lower redshifts ($z\lesssim 9$), when the completely ionized segment ($\ehi=0$) percolates, is not quite interesting from statistical point of view. Therefore, we exclude analysis of shape distribution of HI underdense clusters at lower redshifts, $z\lesssim 9$. 
 
\begin{figure*}
\centering
\subfigure[distribution of cross-section]{
\includegraphics[width=0.46\textwidth]{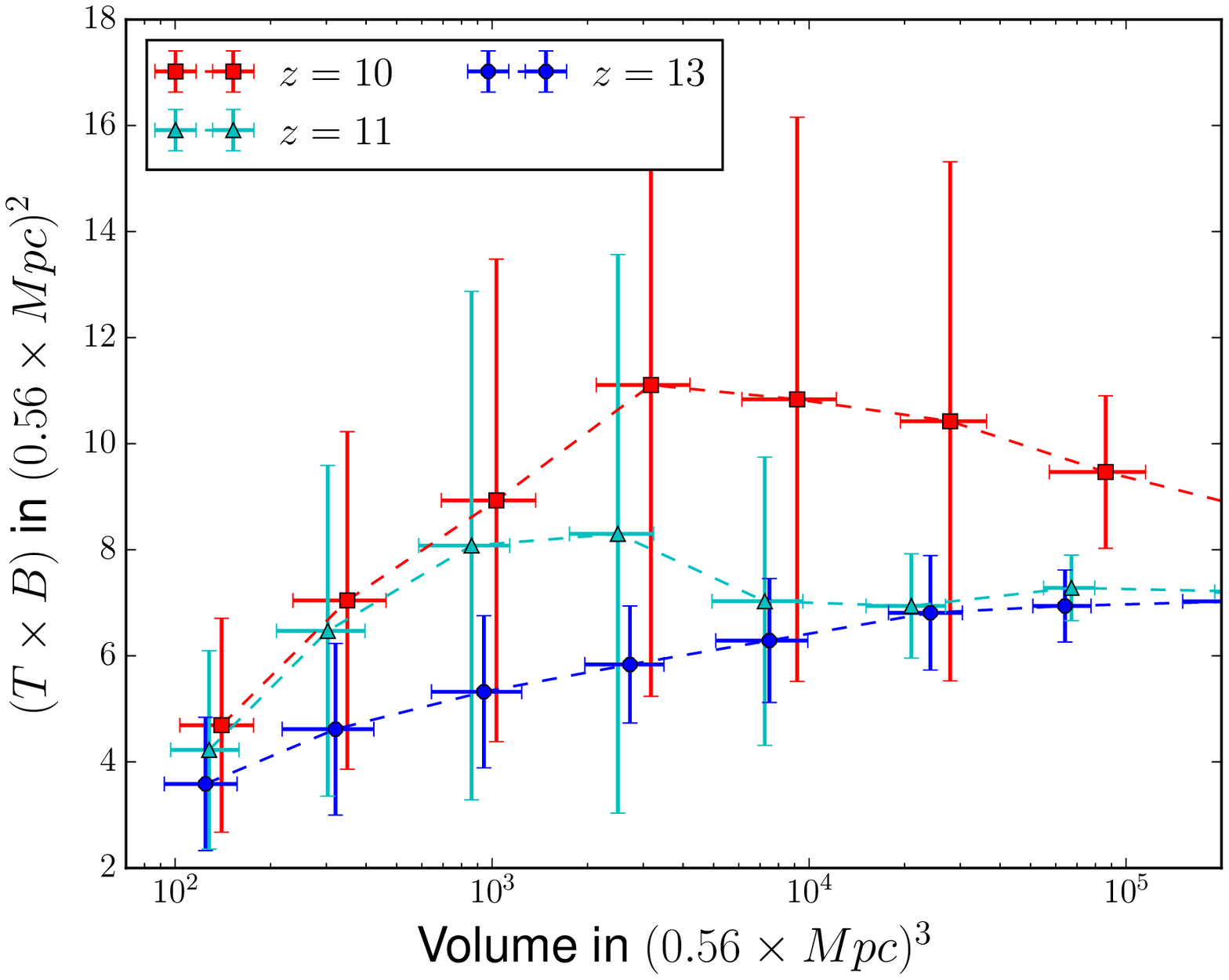} \label{fig:under_Cross_TB}}
\subfigure[evolution of slopes]{
\includegraphics[width=0.49\textwidth]{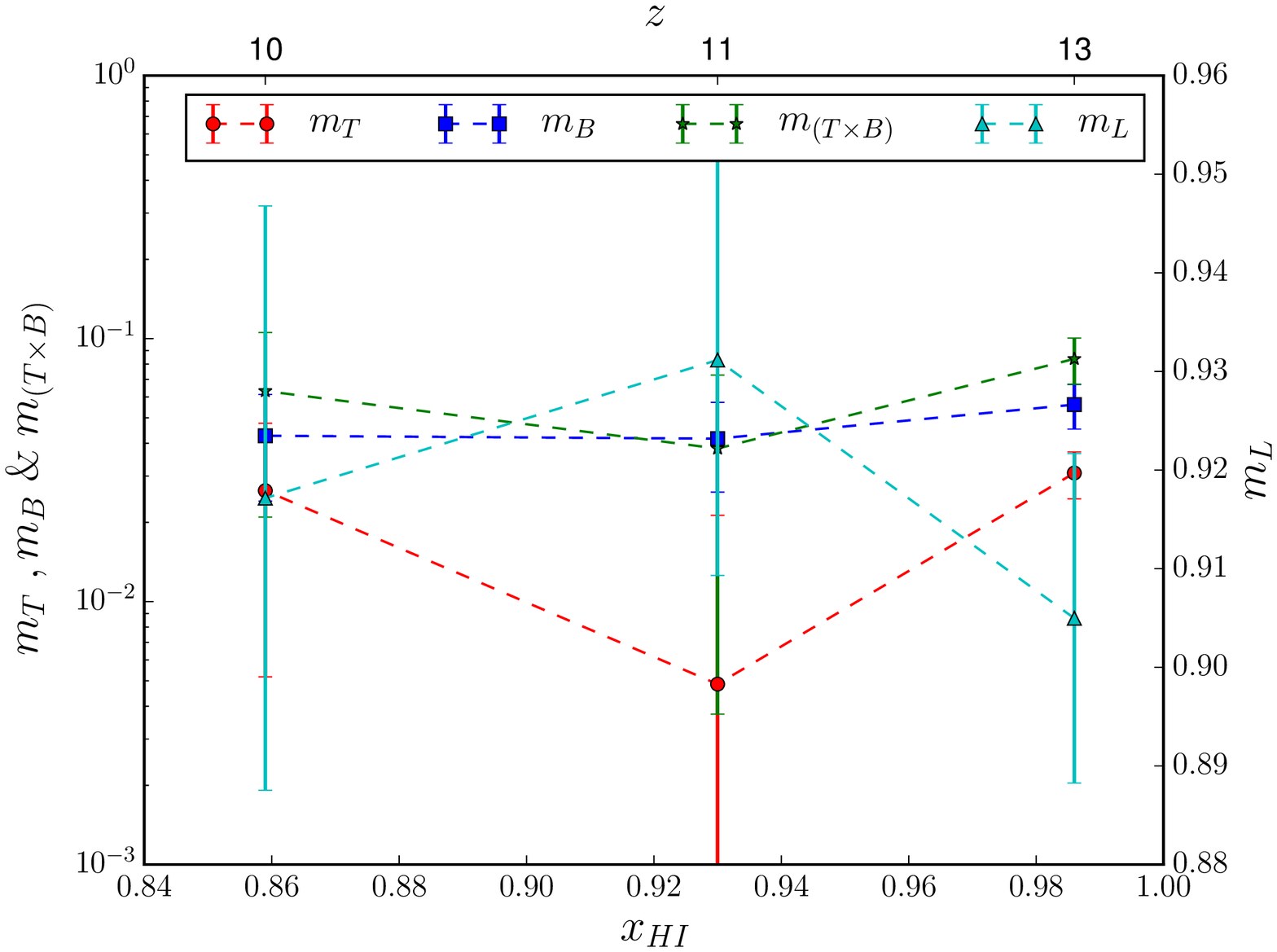}\label{fig:under_m_TBL}}
\caption{  {\bf (a)} In the left panel, we plot $T \times B$, which can be interpreted as the `cross-section' of a filamentary cluster, in volume bins just before percolation in the HI underdense segment at various redshifts. Since, for $z \lesssim 9$, the underdense segment percolates for any $\eth >0$, we exclude the shape analysis of underdense segment at these lower redshifts. {\bf(b)} The slopes of the best fit straight lines to Shapefinders of HI underdense clusters are plotted for various neutral fractions. The slopes are defined as $m_X \equiv \dd \log X/\dd \log V$ where $X \in (T,B, T\times B, L)$. The slopes $m_T$, $m_B$, $m_{T \times B}$ are plotted along left y-axis in log scale while $m_L$ is plotted along right y-axis in linear scale. }
\label{fig:void_TBL_slopes}
\end{figure*}

 For all the redshifts in the range $13 \gtrsim z \gtrsim 9$, we find that the filamentarity increases with cluster volume and saturates to almost unity for larger clusters. Planarity again is very low and increases very slowly with cluster volume. In figure \ref{fig:under_Cross_TB}, the cross-section of underdense clusters (estimated by $T \times B$) is plotted in volume bins for different redshifts\footnote{ For each redshift, again we set the density threshold $\eth = \ehic$.}. We notice that large HI underdense clusters possess somewhat similar cross-section but the cross-sections vary slightly more with cluster volume when compared with that of HI overdense segment (compare figures \ref{fig:under_Cross_TB} with \ref{fig:Cross_TB}).

 We plot the slopes of the best fit straight lines to the Shapefinders, planarity and genus as function of neutral fraction in figure \ref{fig:under_m_TBL}. The redshifts corresponding to the neutral fractions are shown along the top x-axis in both the panels. It is evident that $m_T$, $m_B$, $m_{T \times B}$ for HI underdense segment are slightly higher than that in overdense segment at respective redshifts. This implies that $T$, and especially $B$ of large underdense clusters are less invariant with cluster volume, in comparison with that of the HI overdense segment. On the other hand, the slope of the best fit straight line to  $ \log L$ vs $\log V$ curve is little lower in underdense segment than that in overdense segment. For example, in underdense segment $m_L=0.905 \pm 0.017$ at $z=13$, which is slightly lower than unity when compared with that of the overdense segment at the same redshift.  
 
The following points can be noted regarding Shapefinders for HI underdense clusters:
\begin{itemize}
 \item By extrapolating $P$ and $F$ (e.g. in figure \ref{fig:void_bin0_PF_z13_iso41_1_c100}) towards lower volume at all redshifts, one can conclude that most smaller underdense clusters have very low values of planarity, filamentarity while the genus value is mostly zero. Hence small HI underdense clusters are somewhat like spherical bubbles with trivial topology at all redshifts.
 
 \item Similar to overdense segment, larger clusters in HI underdense segment are statistically very much filamentary. Since maximum number of large clusters appear in the vicinity of percolation transition, again filamentarity becomes overwhelming at these thresholds.
 
 \item Thickness ($T$), breadth ($B$) and the cross-section ($T \times B$) of large underdense clusters are slightly less invariant with cluster volume when compared with that of large overdense clusters, even at the onset of percolation. But they vary with volume in such a way that planarity $P$ is almost invariant, like in the case of overdense segment. But the slope of the best fit straight line for $\log L$ vs $\log V$ curve is not as close to unity as we found in the case of overdense segment.

\end{itemize}

\subsection{Comparison with the hydrogen field}

\begin{figure*}
\centering
\subfigure[Hydrogen overdense segment (structure), $\eth=2.1328$]{
\includegraphics[width=0.483\textwidth]{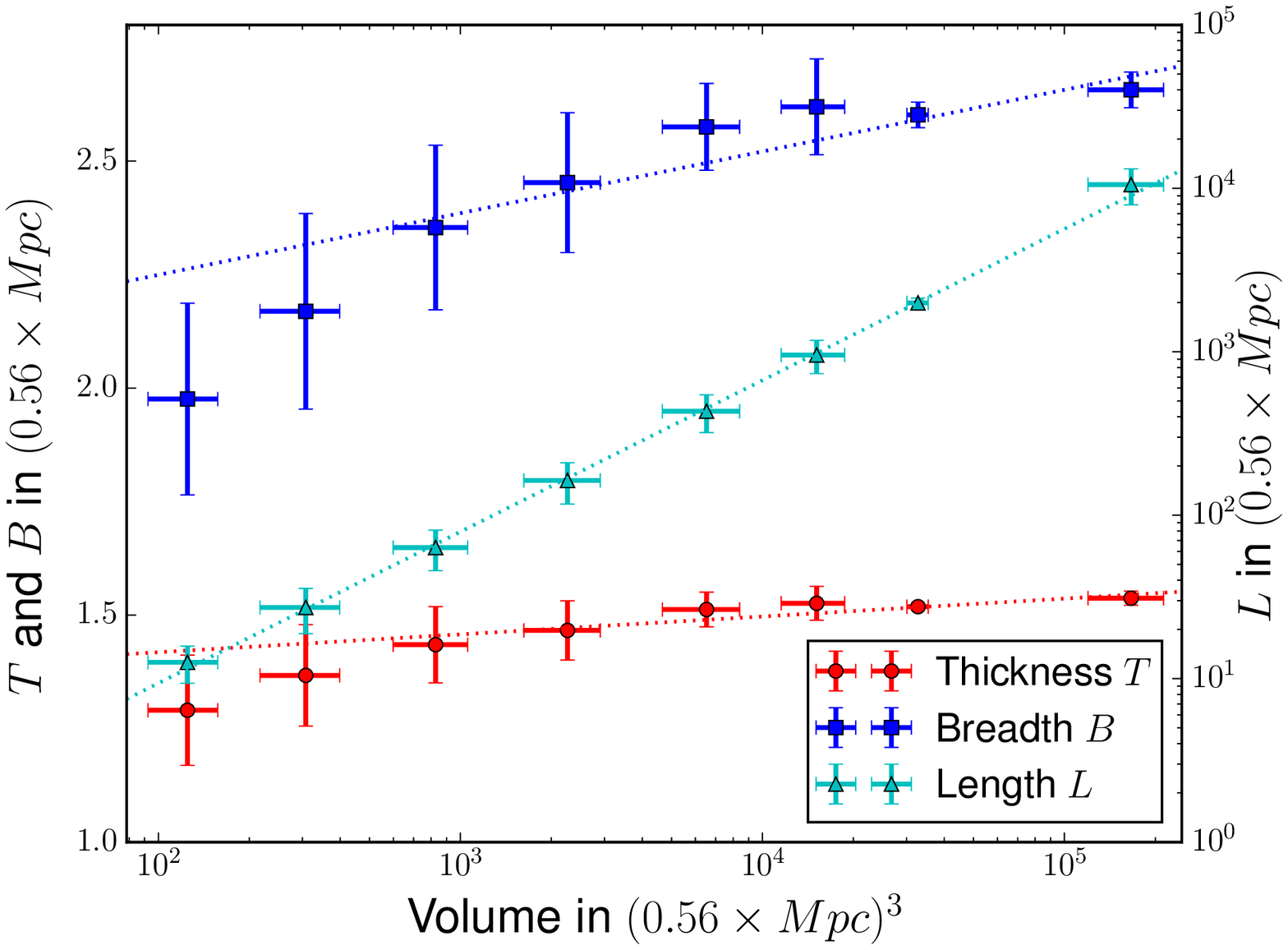}\label{fig:bin0_TBL_z8_struc_Hfield_iso136_5_c100}}
\subfigure[Hydrogen underdense segment (void), $\eth=0.5031$]{
\includegraphics[width=0.483\textwidth]{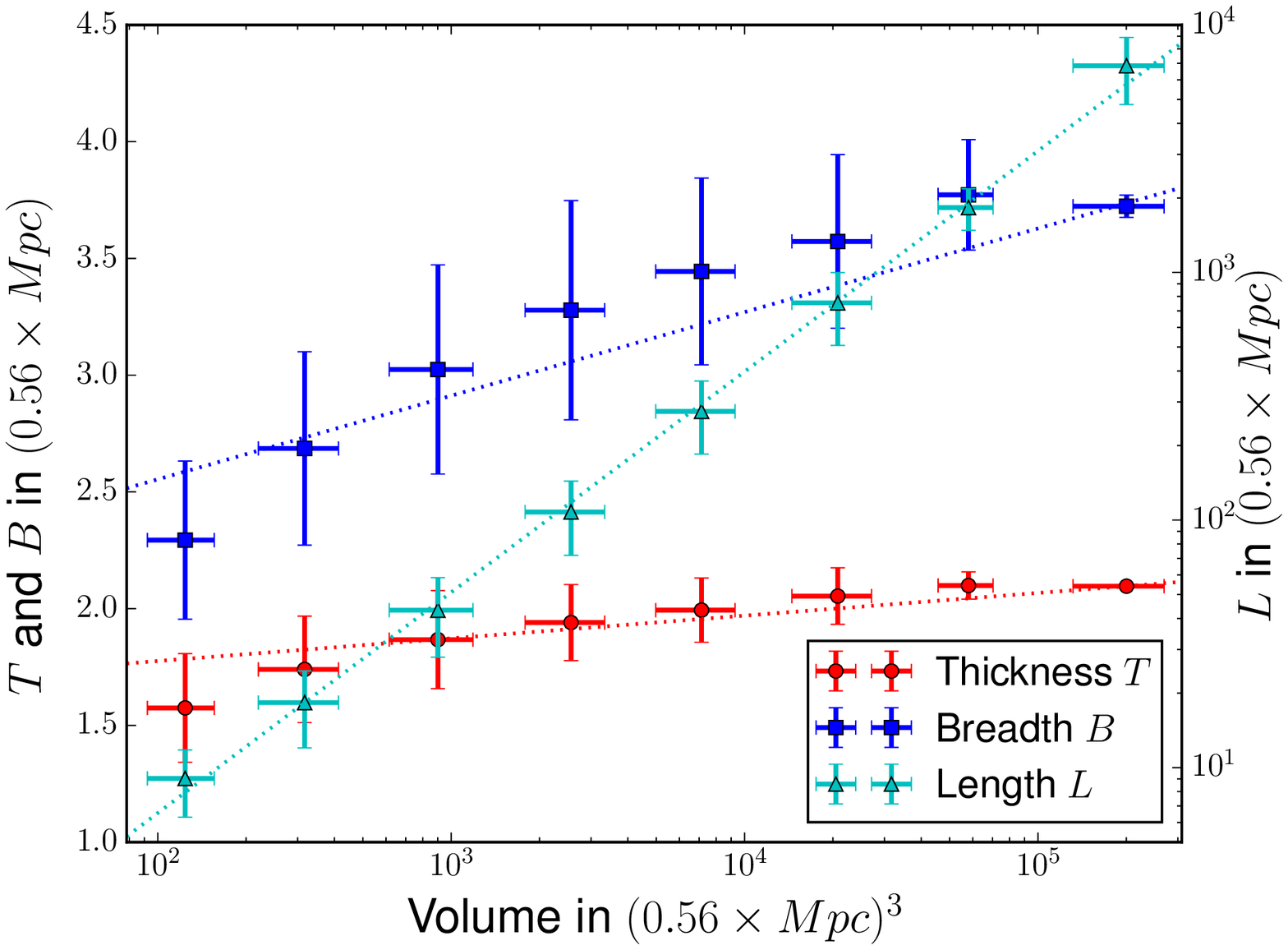}\label{fig:void_bin0_TBL_zH_field_iso32_2_c100}}
\caption{ $T,B,L$ of the clusters in over-density (structure) and under-density (void) excursion sets in hydrogen field are shown at the onset of respective percolation at $z=8$ in left and right panels respectively. The Shapefinders values are shown in suitable volume bins.
 For the overdense clusters, shown in the left panel, the slopes of the best fit straight lines to Shapefinders are given by $m_T=0.023$, $m_B=0.040$, $m_{T \times B}=0.062$ and $m_L=0.932$. The values of $m_T$, $m_B$ and $m_{T \times B}$ are significantly higher than the corresponding values for HI density field at the same redshift. But these slopes for hydrogen field at $z=8$ match well with that of HI field at $z=13$. The plot of Shapefinders in the right panel for void in hydrogen field is completely different to that of HI field at the same redshift. Rather $T, ~B, ~L$ for hydrogen field at $z=8$ seem to vary with volume similarly  as the HI field at high redshift ($z=13$), for example compare figure \ref{fig:bin0_TBL_z8_struc_Hfield_iso136_5_c100} and \ref{fig:void_bin0_TBL_z13_iso41_1_c100}.}
\label{fig:Hfield}
\end{figure*}

It would be interesting to compare the morphology of excursion sets in HI density field with that of the total hydrogen density field. The comparison can highlight the crucial differences in morphology of clusters in hydrogen and HI field and therefore may enlighten the effect of reionization.  
At $z=8$, the overdense (structure) and underdense (void) excursion sets in the total hydrogen field, defined according to equation \eqref{eq:segments}, percolate at the critical thresholds $\eth=2.1328$ ($\ffc=0.050$) and $\eth=0.5031$ ($\ffc=0.127$) respectively. Both critical thresholds are significantly different from the corresponding critical values for HI field at the same redshift, see table \ref{table:percolation}.
In figure \ref{fig:Hfield} we plot the Shapefinders of overdense (left panel) and underdense (right panel) clusters in hydrogen field, again just before respective percolation transitions, at $z=8$. Similar to HI field, the thickness ($T$) and breadth ($B$) of large clusters of hydrogen overdense region (structure) do not increase much with volume. But $T$ and $B$ of the hydrogen overdense clusters grow slightly more with increasing volume than that of the HI field at same redshift $z=8$, compare  figures \ref{fig:bin0_TBL_z8_struc_Hfield_iso136_5_c100} and \ref{fig:bin0_TBL_z8_iso68_5_c100}. Actually the shape distribution in figure \ref{fig:bin0_TBL_z8_struc_Hfield_iso136_5_c100} resembles the corresponding plots for HI field at higher redshift, when the neutral fraction is very close to unity (for example at $z=13$; compare  figures \ref{fig:bin0_TBL_z8_struc_Hfield_iso136_5_c100} and \ref{fig:bin0_TBL_z13_iso101_5_c100}).
The Shapefinders $T,B,L$ for underdense clusters in total hydrogen density field is shown in figure \ref{fig:void_bin0_TBL_zH_field_iso32_2_c100} just before percolation at $\eth=0.5031$.  Note that at $z=8$, in contrast to hydrogen field, the underdense segment in HI density field percolates for nonzero any threshold, $\eth>0$. Again the Shapefinders plotted in figure \ref{fig:void_bin0_TBL_zH_field_iso32_2_c100} looks similar to the corresponding plots at $z=13$ (compare with the figure \ref{fig:void_bin0_TBL_z13_iso41_1_c100}).

In summary, we observe that the morphology of the total hydrogen field at $z=8$ is different from that of the HI field at the same redshift (rather the former is quite similar to that of HI field at high redshifts when the neutral fraction is almost unity \footnote{At high redshifts, the morphology of HI field is expected to be similar to that of the hydrogen field because the IGM is mostly neutral. However, from the observational perspective, the 21-cm signal is more complex due to spin temperature at high redshifts. For simplicity, we always assume $T_s \gg T_{\gamma}$ in the simulations.}). Therefore, the feature that the similarity of $T$, as well as $B$, of large HI overdense clusters is enhanced at lower redshifts (as explained in sections \ref{subsec:over}) is not a generic characteristic of the hydrogen field. The contrast in morphology of HI field and hydrogen field at the same redshift confirms that the above feature has certainly appeared in the HI field due to the reionization of the neutral hydrogen.

\subsection{Volume averaged Shapefinders as a function of \texorpdfstring{$\eth$}{Lg}}

\begin{figure*}
\centering
\subfigure[$z=7$]{
\includegraphics[width=0.483\textwidth]{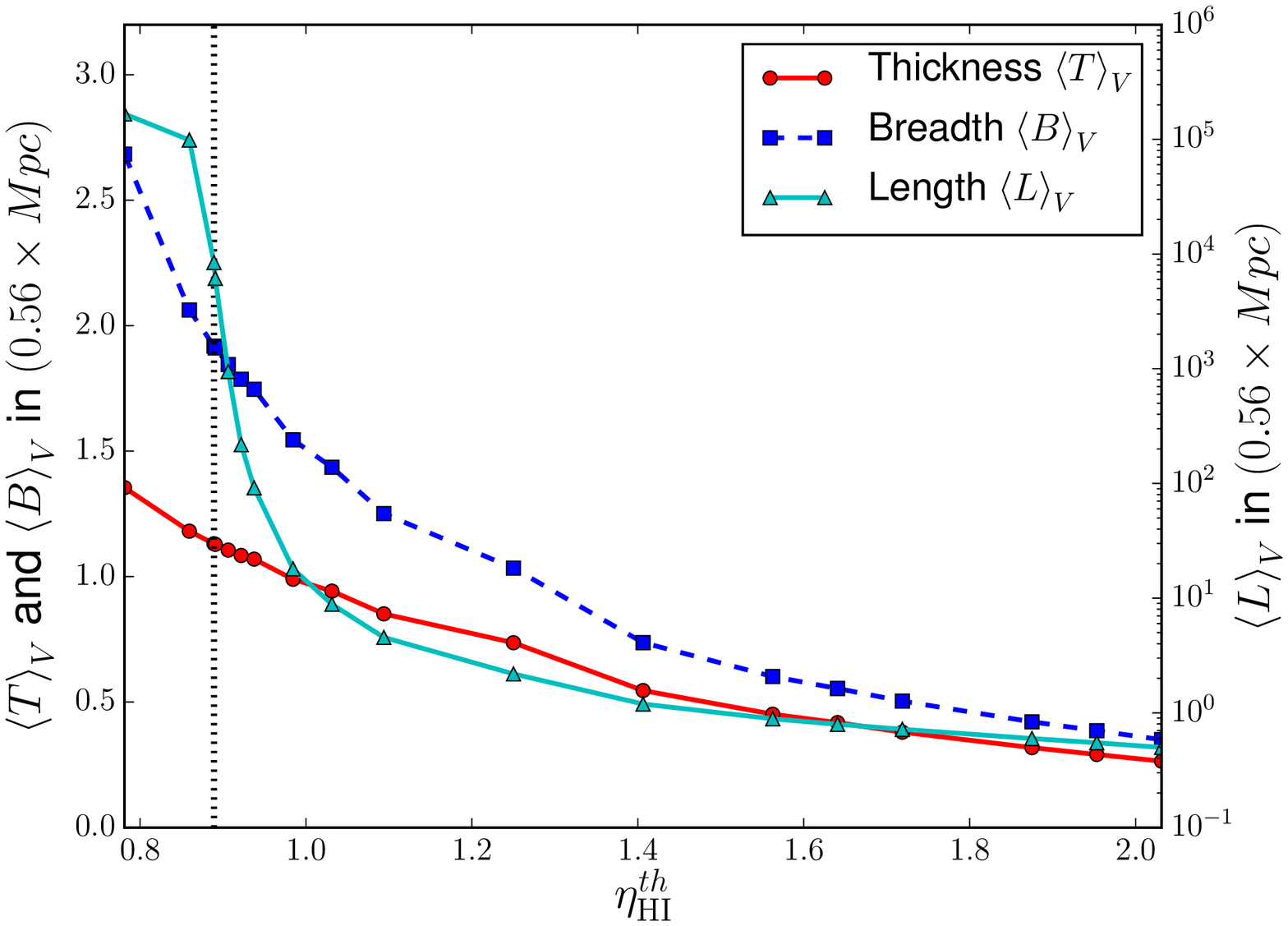}\label{fig:TBL_delth_z7_new}}
\subfigure[Comparison between $z=7$ and $z=13$ ]{
\includegraphics[width=0.483\textwidth]{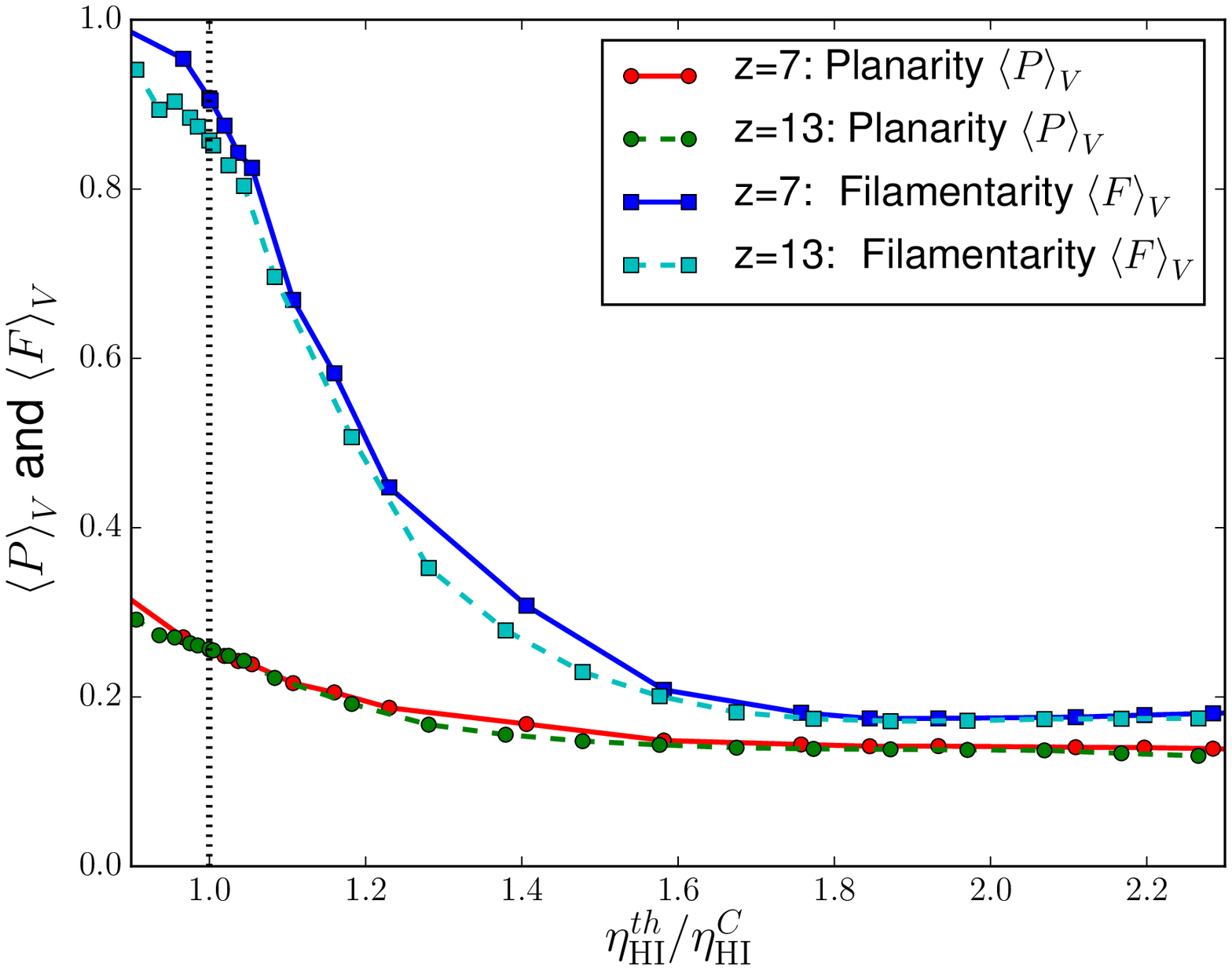}\label{fig:PF_delth_z7_new_comp}}
\caption{ {\bf (a):} The volume averaged thickness ${\langle T \rangle}_V$, breadth ${\langle B \rangle}_V$ and length ${\langle L \rangle}_V$ of all overdense clusters are plotted as a function of the threshold $\eth$ at $z=7$. Different scales have been used to plot  ${\langle T \rangle}_V$ and ${\langle B \rangle}_V$ (along left y-axis) on the one hand and ${\langle L \rangle}_V$ (along right y-axis) on the other.  {\bf (b):} The plots of volume averaged planarity and filamentarity of overdense clusters against $\eth$ are compared for two redshifts $z=7$ and $13$. The threshold along x-axis is normalized by the respective critical threshold at percolation at these redshifts. The evolution of ${\langle P \rangle}_V$ and ${\langle F \rangle}_V$ for the two redshifts are very much alike. Note that the HI overdense segment is percolating below the critical density threshold shown by the black dotted line in both the panels.}
\label{fig:struc_SFv_z7}
\end{figure*}

\begin{figure*}
\centering
\subfigure[$z=10$]{
\includegraphics[width=0.465\textwidth]{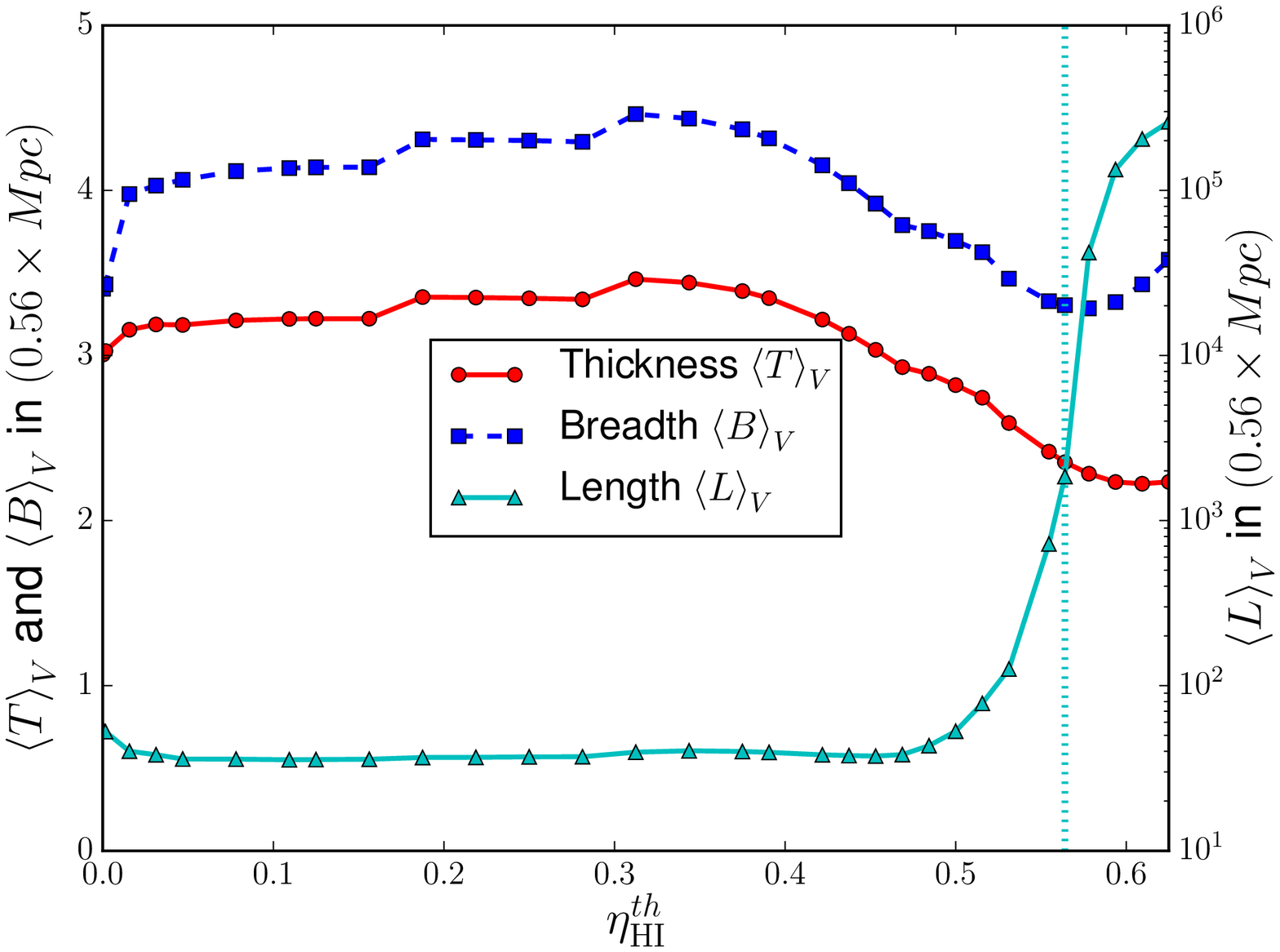}\label{fig:TBL_void_delth_z10_new}}
\subfigure[Comparison between $z=10$ and $z=13$]{
\includegraphics[width=0.49\textwidth]{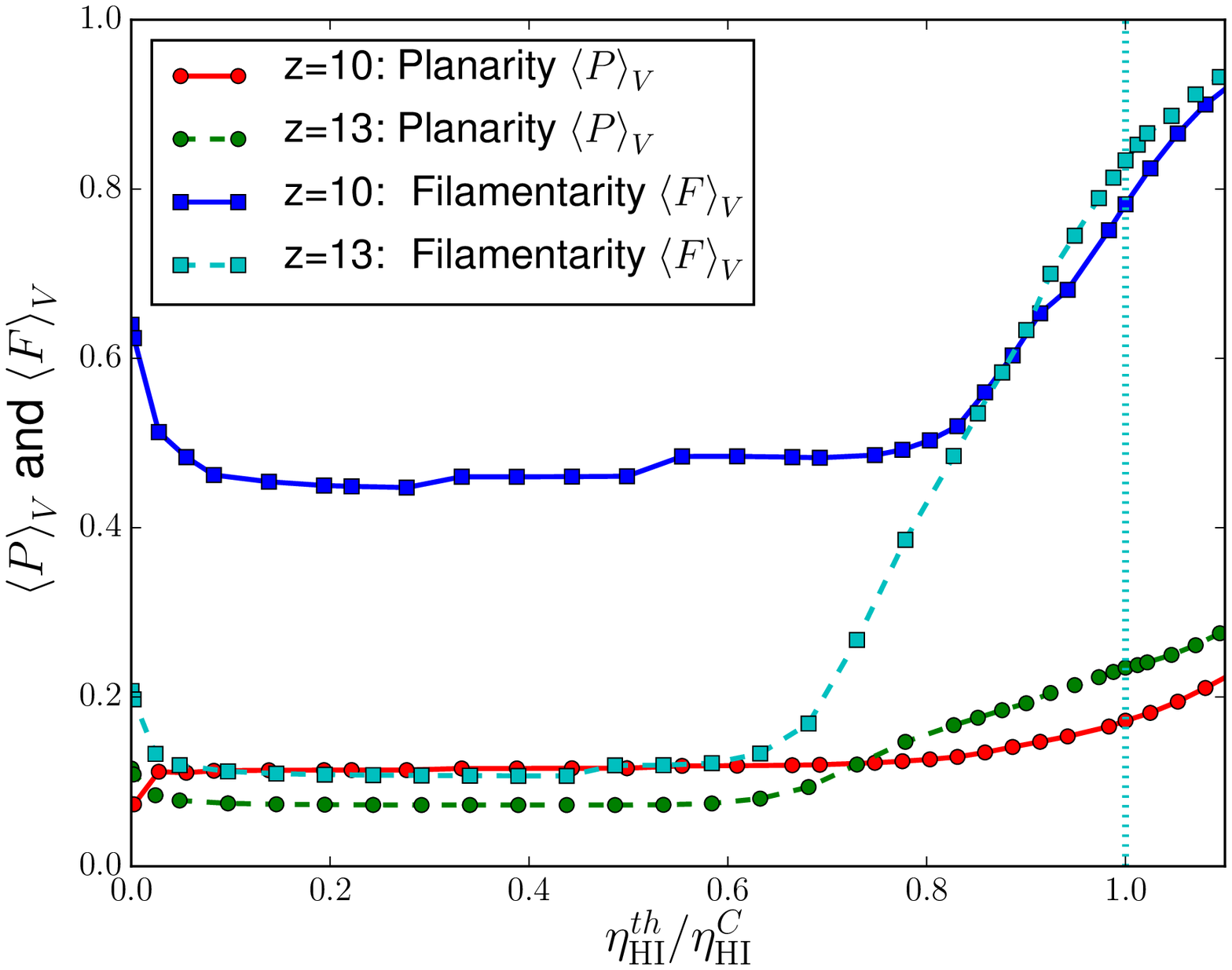}\label{fig:SF_void_rth_z10_comp}}
\caption{ {\bf (a):} The volume averaged Shapefinders of the clusters in HI underdense segment at $z=10$ are plotted as a function of $\eth$. Note that linear scale has been used to plot ${\langle T\rangle}_V$ and  ${\langle B \rangle}_V$ along the left-y axis while ${\langle L \rangle}_V$ has been plotted along right y-axis in log scale.
{\bf (b):} Volume averaged planarity and filamentarity of HI underdense clusters segment as a function of density threshold are compared for two redshifts, $z=10$ and $13$. $\eth$ along x-axis is normalized by the critical threshold at percolation transition at the two redshifts.
}
\label{fig:void_SFv}
\end{figure*}

To understand the global morphology we study Shapefinders, volume weighted averaged over all the clusters, as a function of the density threshold $\eth$ in both HI overdense and underdense excursion sets. In figure \ref{fig:TBL_delth_z7_new}, the volume averaged  Shapefinders of all overdense clusters at $z=7$ are plotted against $\eth$. The percolation transition is shown by the black vertical dotted line. For a high value of density threshold, there are very few small overdense clusters which are expected to be mostly spherical; hence volume averaged thickness (${\langle T\rangle}_V$), breadth (${\langle B \rangle}_V$) and length (${\langle L \rangle}_V$) are quite small for high values of $\eth$ and have same order of magnitude. Note that linear scale has been used to plot ${\langle T\rangle}_V$ and  ${\langle B \rangle}_V$ along the left-y axis while ${\langle L \rangle}_V$ is plotted along right y-axis in log scale. As the threshold is lowered, more large clusters start to appear, and their contribution to  volume averaged quantities increases. Hence ${\langle T\rangle}_V$, ${\langle B \rangle}_V$ and  ${\langle L \rangle}_V$ increase with decreasing $\eth$. But at the onset of percolation, ${\langle L \rangle}_V$ increases much more rapidly (in $\log$ scale) compared to that of ${\langle T\rangle}_V$ and ${\langle B \rangle}_V$, as shown in figure \ref{fig:TBL_delth_z7_new} for $z=7$. That is because, at the vicinity of percolation many large filamentary clusters appear and dominate the contribution to the volume averaged Shapefinders. The volume averaged planarity ${\langle P \rangle}_V$ and filamentarity ${\langle F \rangle}_V$ in HI overdense segment are plotted against density threshold for two redshifts $z=7$ and $13$ in figure \ref{fig:PF_delth_z7_new_comp}. As expected, ${\langle F \rangle}_V$ increases much more sharply than ${\langle P \rangle}_V$ at the onset of percolation at each redshift.  
Note that, along x-axis in figure \ref{fig:PF_delth_z7_new_comp}, $\eth$ is normalized by the corresponding critical density threshold at percolation ($\ehic$) for each redshift. It is evident that ${\langle P \rangle}_V$ and ${\langle F \rangle}_V$ at the redshifts $z=7$ and $13$ grow quite similarly with decreasing $\eth$ (when normalized by  $\ehic$).
The largest clusters well after percolation encloses most of the volume but its Shapefinders can not be calculated precisely. Hence we do not plot volume averaged Shapefinders far beyond the percolation transition in this section.

In figure \ref{fig:TBL_void_delth_z10_new}, volume averaged Shapefinders for the HI underdense segment are plotted against $\eth$ at $z=10$.  Even at very low $\eth$, the volume averaged Shapefinders are quite high because of the contributions from the large completely ionized clusters ($\ehi=0$) which enclose significant fraction of the total volume. As $\eth$ is increased from a very low value, ${\langle L \rangle}_V$ first decreases and then almost saturates. During the percolation transition in the underdense segment, shown by the vertical dotted cyan line, ${\langle L \rangle}_V$ increases much more sharply than  ${\langle T\rangle}_V$ and ${\langle B \rangle}_V$, similar to what we observe for overdense segment, shown in figure \ref{fig:TBL_delth_z7_new}.  In contrast to overdense segment,  ${\langle T\rangle}_V$ and ${\langle B \rangle}_V$ for underdense segment do not increase monotonically and remain stable as $\eth$ approaches the critical value at percolation transition.

In figure \ref{fig:SF_void_rth_z10_comp}, the volume averaged planarity and filamentarity of the HI underdense segment are compared for two redshifts $z=10$ and $13$. Again $\eth$ along x-axis is normalized by the corresponding critical thresholds at percolation transitions for these two redshifts. Unlike the overdense segment, ${\langle P \rangle}_V$, as well as ${\langle F \rangle}_V$, for the underdense segment behaves differently with  $\eth$ at the two redshifts, specially well below the percolation threshold. At a smaller redshift, the completely ionized regions are larger and have higher filamentarity and planarity. Therefore, the volume averaged ${\langle F \rangle}_V$ and ${\langle P \rangle}_V$ for low $\eth$ are higher at $z=10$ than that at $z=13$, as shown in figure \ref{fig:SF_void_rth_z10_comp}. Again at the onset of percolation transition, shown by the cyan dotted vertical line, ${\langle F \rangle}_V$ increases more steeply than ${\langle P \rangle}_V$.

\section{Conclusion and discussions} \label{sec:conclusion}

Minkowski functionals and Shapefinders are powerful statistical tools to study the clustering properties of large scale cosmological fields from a geometrical perspective. They can act as a complement to the $N$-point correlation functions. In our preceding work \citep{Bag:2018zon}, we found that the completely ionized region starts to percolate reasonably early in the reionization history. For the default reionization model, the percolation transition takes place in the ionized segment at $z\approx 9$ or equivalently $\xhi \approx 0.728$. The study of Shapefinders reveals that the largest ionized region becomes more filamentary as the density threshold approaches the percolation transition.

As a continuation to our previous work, in this work, we have studied the morphology of the HI field, from the excursion set approach, near the epoch of reionization using semi-numerical simulations \citep{Choudhury2009}. \footnote{ For simplicity in the simulations, we assume that CMB temperature is negligible in comparison with the spin temperature, i.e. $T_{\gamma} \ll T_s$. However, the 21-cm signal is more complicated due to the spin temperature at high redshifts. We also set the bias to unity implying that the hydrogen density exactly traces the underlying dark matter distribution.} In particular, we have explored the ``clusters'' in HI overdense and underdense excursion sets in the simulated HI field, where the separation of the over- and underdense regions are carried out in terms of a threshold $\eth$ in the HI density field. The default reionization history has been chosen so as to achieve $\sim 50\%$ global ionization fraction at $z = 8$. The main results of this work are as follows:
\begin{itemize}


\item The overdense and underdense segments percolate at different critical density thresholds corresponding to different values of the respective filling factors. To explore this asymmetry in percolation more quantitatively, one can study the so called ``percolation curves'' which are essentially plots of the filling factors of the largest cluster against the corresponding $\ffc$ for overdense and underdense segments. The area under the percolation curve is a robust geometric measure of non-gaussianity \citep{Sahni:1996mb}. We find that the area under the hysteresis in the percolation curves at $z=13,11$ and $10$ increases as reionization proceeds which in turn indicates that the non-gaussianity in the HI density field increases.

\item Since most of the large clusters appear just before percolation, we have studied the behaviour of the Minkowski functionals for values of density thresholds corresponding to the onset of respective percolation transitions in both overdense and underdense segments. The clusters have different Minkowski functionals but their ratios, defined as Shapefinders, show interesting properties. 

The thickness ($T$) and breadth ($B$) of large clusters in both overdense and underdense segments are of similar values but the length ($L$) increases almost linearly with the volume of the cluster. Hence the filamentarity ($F$) of large clusters in both overdense and underdense segments increases with their volume and saturates to unity for extremely large clusters. The similar values of $T$ and $B$ imply that the filament-like large clusters have similar cross sections (estimated by the product of the first two Shapefinders $T \times B$) and their shapes only differ in terms of their lengths. The reason is that the larger clusters form due to mergers of relatively smaller clusters (as the density threshold is lowered (raised) for overdense (underdense) segment) which are themselves large enough to possess similar values of $T$ and $B$.

The genus of the cluster too increases with the volume. A high value of genus makes the clusters porous with many tunnels passing through. One can thus imagine the large clusters have multiply connected structures made of many filamentary branches and sub-branches. These filamentary sub-structures across large clusters possess similar (characteristic) cross sections, estimated by $T \times B$. The third Shapefinder length $L$ provides some measure of the total length of these filamentary branches which interconnect to form the cluster. At a fixed redshift, as we lower (raise) the density threshold for overdense (underdense) segment, more such filamentary branches (with the characteristic cross-section) connect to the clusters resulting in increase of $L$ proportional to the growth of volume of the clusters, while $T$ and $B$ remain almost constant.  

Moreover the $T$ and $B$ of any large cluster are of comparable values which result in quite low planarity $P$ which is also similar across all large clusters.

The similarity of $T$ as well as $B$ among large clusters at the onset of percolation seems to be more profound at later stages of reionization (at lower redshifts) than that at higher redshifts. Although this feature is quite generic but certainly occurs more vividly in the overdense segment compared to the underdense segment. 

\item For comparison we have also studied the  Shapefinders of clusters (in both overdense and underdense excursion sets) in the baryonic density field. We find that the behaviour of the thickness $T$ and breadth $B$ for the large overdense clusters in HI density field at lower redshifts is not a generic feature of the baryon density field, it is rather driven by the growth of ionization bubbles.

\item For a measure of global Shapefinders, we have also studied the average of Shapefinders of all clusters weighted by the cluster volume at different $\eth$. For both HI overdense and underdense segments, ${\langle L \rangle}_V$ increases steeply at the onset of percolation but  ${\langle T \rangle}_V$ and ${\langle B \rangle}_V$ do not rise that sharply. Also the volume averaged filamentarity ${\langle F \rangle}_V$ increases very sharply but not ${\langle P \rangle}_V$. We find that the increase in ${\langle F \rangle}_V$ and  ${\langle P \rangle}_V$ in overdense segment is very much similar for the two redshifts $z=7$ and $13$ when plotted against $\eth$ normalized by the critical threshold at percolation.

\end{itemize}

Our results complement the recent findings of \cite{Yoshiura:2014ria}, who utilized the global Minkowski functionals to study the topology and morphology of the 21-cm line brightness-temperature fields during reionization. They studied the time-evolution of global Minkowski functionals, which also demonstrates the increase of non-Gaussianity qualitatively in the brightness-temperature fields, before and during the epoch of reionization. On the other hand, we studied the distribution of shapes of clusters in both HI overdensity and underdensity excursion sets at various stages of reionization. This directly reveals the evolution of the shape distributions as reionization progresses. Additionally, we estimate the amount of non-Gaussianity from geometrical point of view using the `percolation curves'.

In continuation of our previous work \citep{Bag:2018zon}, this work is the second in a series of papers to explore the shape statistics of the reionization field. The statistics used here contain information that are not present in the conventional probes of reionization, e.g., two-point correlation function. It would be interesting to explore which of the physical processes related to reionization can be probed using the Shapefinders. This would require computing the statistics for a variety of reionization models involving different physics. We also plan to extend our analysis to understand whether the data from the upcoming low-frequency interferometers can be used for calculating the Shapefinders. For this exercise, we have to compute Shapefinders in the presence of instrument noise and astrophysical foregrounds.


\section*{Acknowledgement}
The authors would like to acknowledge the crucial inputs from Santanu Das, Aseem Paranjape and Ajay Vibhute in developing the code SURFGEN2. S. B. thanks the Council of Scientific and Industrial Research (CSIR), India, for financial support as senior research fellow. V.S. is partially supported by the J.C.Bose Fellowship of
Department of Science and Technology, Government of India. The HI simulations, used in this work, were done at the computational facilities at the Centre for Theoretical Studies, IIT Kharagpur, India. The numerical computations, related to shape analyses, were carried out using the high performance computation (HPC) facilities at IUCAA, Pune, India.

\appendix

\section{Calculating Integrated mean curvature on a triangulated surface}\label{appendix:imc}
\begin{figure}
\centering
\includegraphics[width=0.5\textwidth]{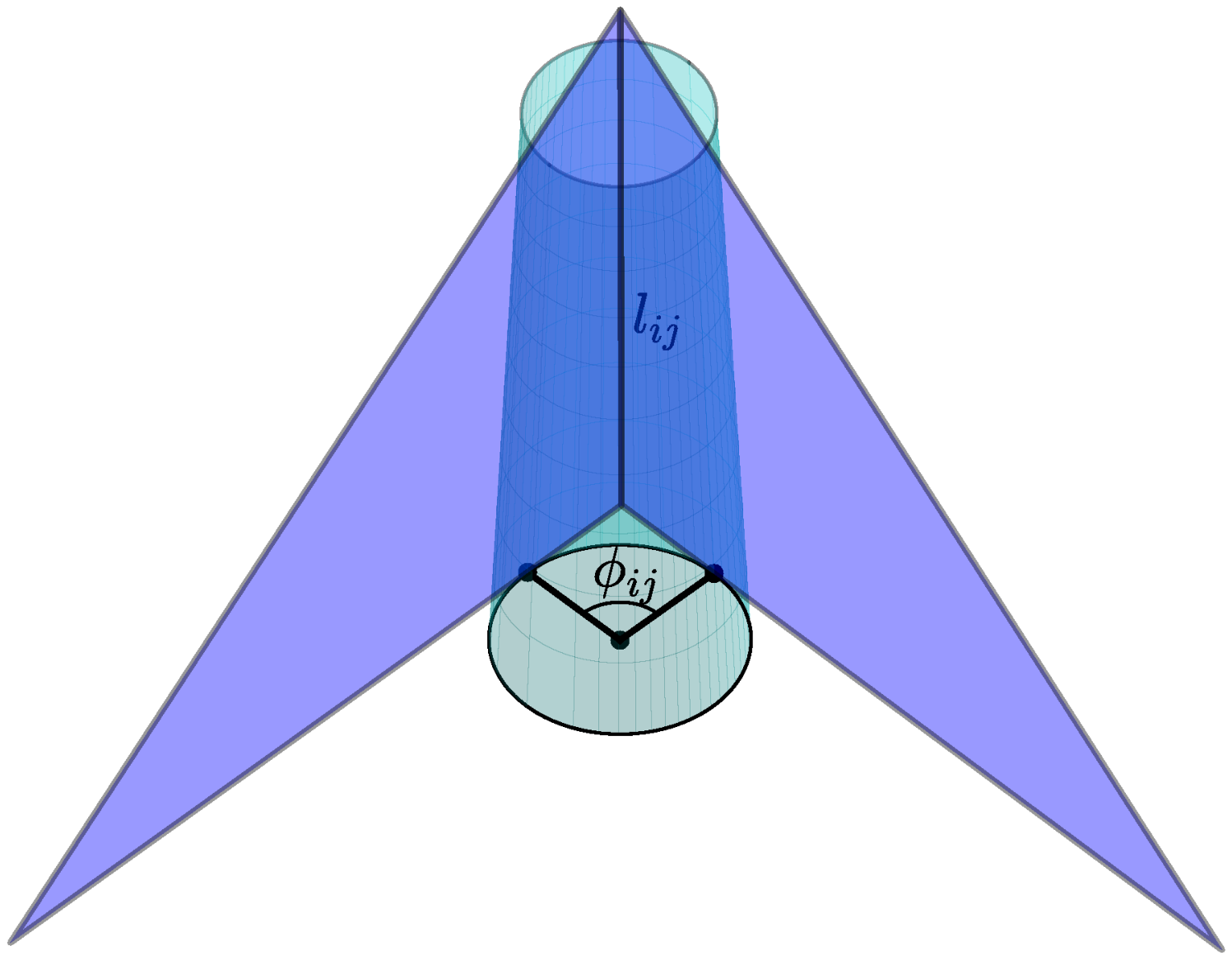}
\caption{The sharp edge, contributing to the integrated mean curvature (IMC), can be regularized by fitting a cylindrical surface to the planes of the adjacent triangles along the edge. Irrespective of the radius of the fitted surface, the projected angle at the centre is $\phi_{ij}$ and the local IMC is $\frac{1}{2} l_{ij} \phi_{ij}$.}
\label{fig:edge_cylinder}
\end{figure}

\begin{figure*}
\centering
\subfigure[locally convex: $\epsilon=+1$]{
\includegraphics[width=0.483\textwidth]{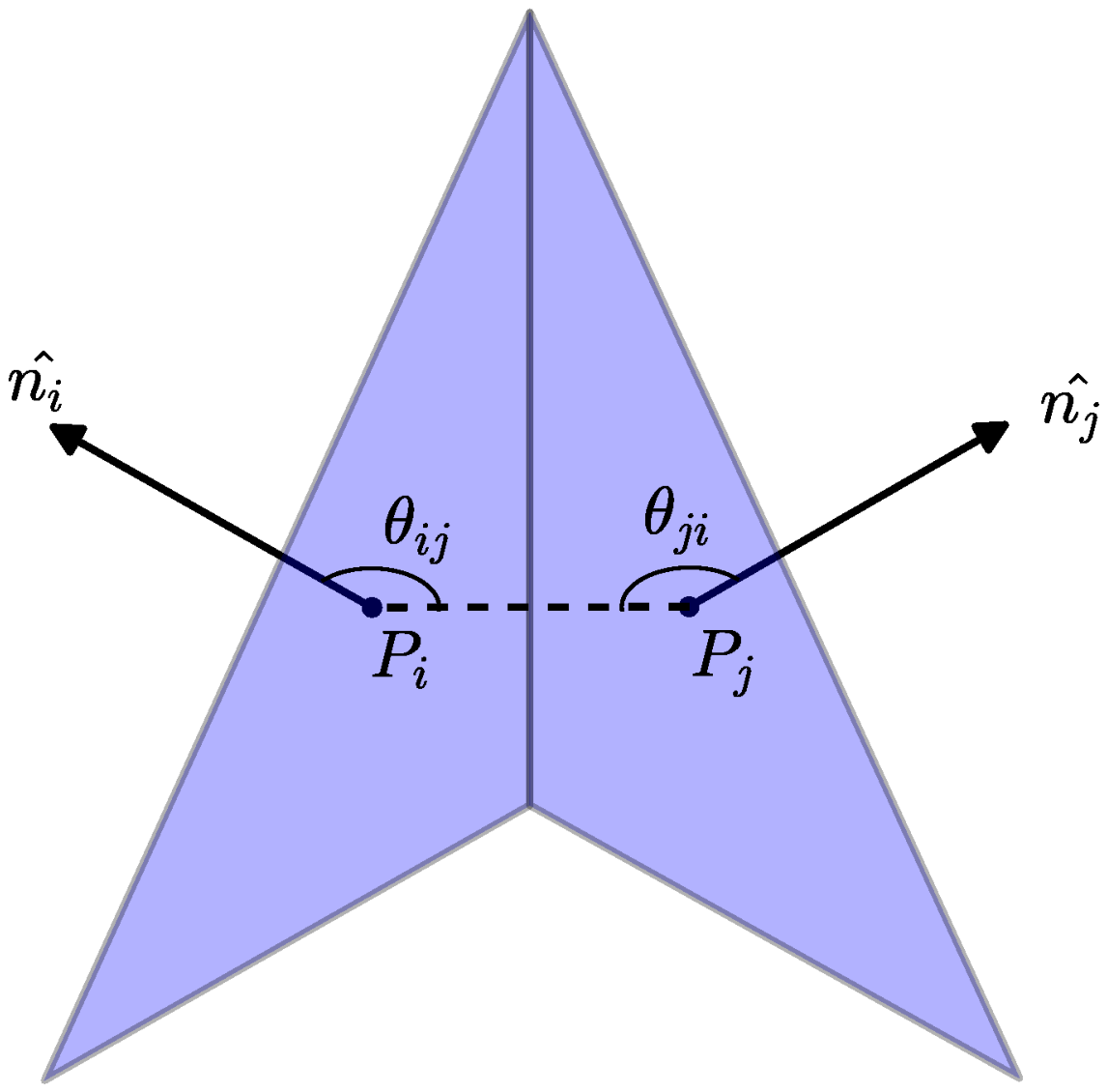}\label{fig:convex}}
\subfigure[locally concave: $\epsilon=-1$]{
\includegraphics[width=0.483\textwidth]{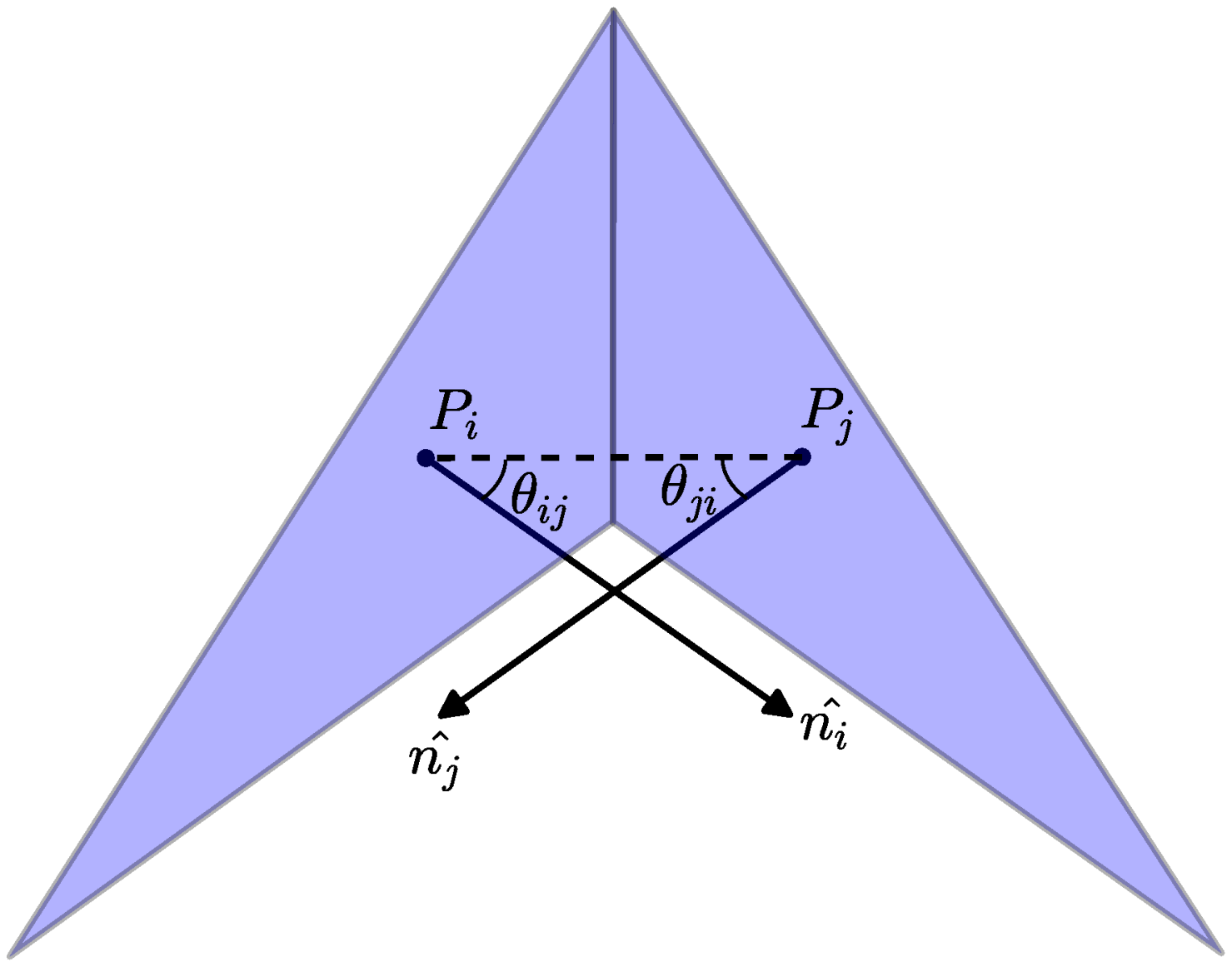}\label{fig:concave}}
\caption{The method of determining the value of $\epsilon$ appearing in equation \eqref{eq:imc} for locally convex (left panel) and concave (right panel) surfaces. $P_i$ and $P_j$ are two arbitrary points  on the triangles. 
The locally convex surface,  has positive contribution to the total integrated mean curvature, hence $\epsilon=+1$. The normal of one triangle will project an obtuse angle to the vector drawn from that triangle to any points on the other triangle, as demonstrated in the left panel. On the other hand for locally concave surface $\epsilon=-1$ because of its negative contribution to total  integrated mean curvature. In that case the normal of one triangle will project an acute angle to the vector drawn from that triangle to the other one, as shown in the right panel. Hence by computing the angle $\theta_{ij}$ one can easily determine whether the surface is locally convex or concave. Note that if $\theta_{ij}$ is acute (obtuse), $\theta_{ji}$ will be also acute (obtuse) and vice versa.}
\label{fig:epsilon}
\end{figure*}

Since the triangles are flat, the integrated mean curvature (IMC) for a triangulated surface is localized only in the triangle edges. For a closed surface triangulated using Marching Cube 33 algorithm, each triangle edge is shared by only two adjacent triangles. The sharp edge can be approximated by fitting a cylindrical surface of small radius $r$ to the planes of the adjacent triangles along the edge as shown in figure \ref{fig:edge_cylinder}. Only contribution to IMC comes from the part of the cylindrical surface projecting an angle $\phi_{ij}$ at the centre, having a total area $r l_{ij} \phi_{ij}$ and constant principle curvatures, $K_1=1/r$ and $K_2=0$. From the equation \eqref{eq:imc}, one can easily estimate the local IMC as, 
\begin{equation}\label{eq:imc_tri}
C_{ij}=\frac{1}{2}\oint \frac{1}{r} dS= \frac{1}{2r} r l_{ij} \phi_{ij}= \frac{1}{2} l_{ij} \phi_{ij}\;.
\end{equation}
Note that since the projected angle $\phi_{ij}$ remains same with varying radius $r$, the local IMC $C_{ij}$ does not depend on the radius of the fitted cylinder.

The total IMC of the surface would be the sum of the contributions from all the edges:  
 \begin{equation}\label{eq:imc1}
    C=\sum_{i,j}\frac{1}{2} \epsilon\  l_{ij}\phi_{ij}\;,
 \end{equation}
 where the summation  is carried over all the edges or equivalently over all the pairs of adjacent triangles. $\epsilon=1$ for locally convex surface for which the normals to the adjacent triangles (pointing out of the surface)  diverge away from each other, as shown in figure \ref{fig:convex}. On the other hand for the locally concave surface, the normals to the adjacent triangles converge (out of the surface) into each other as shown in figure \ref{fig:concave} and the local contribution to IMC is negative, hence $\epsilon=-1$. To determine whether the triangles are locally convex or concave one can calculate the angle ($\theta_{ij}$) between the normal ($\hat{n_i}$) of a triangle at a point ($P_i$) and a vector from that point to any point on the other triangle (say ${\bf \overrightarrow{P_iP_j}}$). $\theta_{ij} >\pi/2$ for locally convex surface while for locally concave pair of triangles $\theta_{ij} <\pi/2$, as demonstrated in figure \ref{fig:epsilon}. Note that if the angle $\theta_{ij}$ is acute so is $\theta_{ji}$ and vice versa. Therefore, by computing the angle $\theta_{ij}$ one can determine whether the pair of triangles are locally convex or concave and set the correct value of $\epsilon$.

\bibliographystyle{mnras}
\bibliography{thesis_bib}

\bsp	
\label{lastpage}
\end{document}